\documentclass[useAMS,usenatbib]{mn2e}
\usepackage{txfonts}
\usepackage{graphicx} 
\usepackage{rotating}

\title[Generative pulsar timing analysis]{Generative pulsar timing analysis}
\author[L. Lentati et al.]{\parbox{\textwidth}{L. Lentati$^{1}$\thanks{E-mail:
ltl21@cam.ac.uk}, P. Alexander$^{1}$, M. P. Hobson$^{1}$}\vspace{0.4cm}\\ %
$^{1}$Astrophysics Group, Cavendish Laboratory, JJ Thomson Avenue,  Cambridge, CB3 0HE, UK\\
}

\begin{document}

\maketitle

\label{firstpage}

\begin{abstract}
A new Bayesian method for the analysis of folded pulsar timing data is presented that allows for the simultaneous evaluation of evolution in the pulse profile in either frequency or time, along with the timing model and additional stochastic processes such as red spin noise, or dispersion measure variations.  We model the pulse profiles using `shapelets' - a complete ortho-normal set of basis functions that allow us to recreate any physical profile shape.  Any evolution in the profiles can then be described as either an arbitrary number of independent profiles, or using some functional form. We perform simulations to compare this approach with established methods for pulsar timing analysis, and to demonstrate model selection between different evolutionary scenarios using the Bayesian evidence.
The simplicity of our method allows for many possible extensions, such as including models for correlated noise in the pulse profile, or broadening of the pulse profiles due to scattering.  As such, while it is a marked departure from standard pulsar timing analysis methods, it has clear applications for both new and current datasets, such as those from the European Pulsar Timing Array (EPTA) and International Pulsar Timing Array (IPTA).

\end{abstract}


\section{Introduction}

When a pulsar is observed, the individual pulses from a single observing epoch are folded together using some fiducial timing model to form an average pulse profile for that observation.  This averaging can be performed in time alone, or also across some frequency range, after which a cross-correlation is performed between the folded profile and a template or model.  This process allows for the creation of a set of `time of arrivals' (TOAs), one for each folded profile, to which a model for the pulsar can then be fit that characterises its orbital motion, timing properties, such as its orbital frequency and spin down, and any additional timing noise.

While the individual pulses from a pulsar can show a high degree of variability (e.g. \citealt{1981ApJ...249..241H}), the average profile obtained from even a few hundred pulses in a single observing epoch shows remarkable stability (e.g. \citealt{1988MNRAS.234..477L}).  These regimes provide different windows into the environment and mechanisms of the pulse emission, where the individual pulses probe the millisecond fluctuations in the properties of the plasma emission (e.g. \citealt{2007A&A...462..257B}) and the folded profile  provides insight into the statistical properties of the pulsar's magnetic field and the geometry of the emitting region.

It is in large part the stability of this average profile that has led to the exceptional timing precision obtainable with pulsars themselves, with decade long observations of millisecond pulsars in particular providing timing measurements with accuracies similar to atomic clocks (e.g. \citealt{1994ApJ...428..713K,1997A&A...326..924M}).

As the precision with which measurements can be made increases, however, so too must the depth of the analysis performed follow in step.  For example, the average profile template used to form the TOAs for a pulsar is typically obtained by fixing the timing model to some particular set of values, and then folding the data from the individual observations using that model.  In the case that the pulsar exhibits significant timing noise, however, deviations from the timing model will cause smearing in the profile, reducing the precision with which TOAs can be measured.

In addition, it is known that the shape of the averaged profile from a single pulsar can show dramatic evolution as a function of frequency (see e.g. Fig. \ref{figure:realprofiles}).  It is thought that this is a consequence of the spectral index of the radiation emitted by the pulsar changing as a function of the distance of the emitting region from the pulsar's surface. For example,  it has been observed that the pulse profile widens with decreasing frequency, signifying that the emission is emanating from higher in the pulsar's magnetosphere (Thorsett 1991; Mitra $\&$ Rankin 2002).

While evolution on large scales might be easily detectable across different frequency bands, evolution across the band must also be accounted for in order to avoid introducing a systematic bias into the estimated TOAs. In \cite{2013ApJ...762...94D}, for example, while a single template profile is used within each receiving band, additional parameters, known as `jumps', are included in the pulsar timing model that fit a constant offset for each 4MHz frequency channel.  These jumps therefore act as a proxy to any frequency evolution in the pulse profile across a given band, however they provide no insight into the physical origin of the evolution taking place. Such insight can only be robustly obtained by explicitly incorporating a model for the profile, and its evolution, simultaneously with the timing analysis. 

In this paper we present a Bayesian framework to accomplish precisely this goal which we call 'Generative Pulsar Timing Analysis' (henceforth GPTA).  GPTA allows us to
simultaneously fit for the pulsar timing model, stochastic signals such as red `spin' noise and dispersion measure variations, as well as a model for the pulse profile, thereby providing a means of robustly estimating evolution in the profile in either frequency or in time.  This framework uses the timing model as a prior on the arrival time of the pulses, generating model TOAs at which to evaluate the pulse profile and thus accounting for our uncertainty in both the shape of the profile and the timing parameters in the final analysis.

In Section \ref{Section:Bayes} we will describe the basic principles of our Bayesian approach to data analysis, giving a brief overview of how it may be used to perform model selection.  In Section \ref{section:Shapelets} we introduce `shapelets', the basis functions from which we will construct our model for the pulse profile.  In Sections \ref{Section:Like} through to \ref{Section:Generate} we introduce the likelihood that we will use, incorporating the pulsar timing model, stochastic signals and the model for the pulse profile and describe how we generate the model TOAs using the timing model.  In Section \ref{Section:simulations} we demonstrate the efficacy of this method in a series of simulations that include timing noise, and evolution in frequency before finally offering some concluding remarks in Section \ref{Section:Conclusions}.

\section[]{Bayesian Inference}
\label{Section:Bayes}

Our method for performing pulsar timing analysis is built upon the principles of Bayesian inference, which provides a consistent approach to the estimation of a set of parameters $\Theta$ in a model or hypothesis $H$ given the data, $D$.  Bayes' theorem states that:

\begin{equation}
\mathrm{Pr}(\Theta \mid D, H) = \frac{\mathrm{Pr}(D\mid \Theta, H)\mathrm{Pr}(\Theta \mid H)}{\mathrm{Pr}(D \mid H)},
\end{equation}
where $\mathrm{Pr}(\Theta \mid D, H) \equiv \mathrm{Pr}(\Theta)$ is the posterior probability distribution of the parameters,  $\mathrm{Pr}(D\mid \Theta, H) \equiv L(\Theta)$ is the likelihood, $\mathrm{Pr}(\Theta \mid H) \equiv \pi(\Theta)$ is the prior probability distribution, and $\mathrm{Pr}(D \mid H) \equiv Z$ is the Bayesian Evidence.

In parameter estimation, the normalizing evidence factor is usually ignored, since it is independent of the parameters $\Theta$.   Inferences are therefore obtained by taking samples from the (unnormalised) posterior using, for example, standard Markov chain Monte Carlo (MCMC) sampling methods.  

In contrast to parameter estimation, for model selection the evidence takes the central role and is simply the factor required to normalize the posterior over $\Theta$:

\begin{equation}
Z = \int L(\Theta)\pi(\Theta) \mathrm{d}^n\Theta,
\label{eq:Evidence}
\end{equation}
where $n$ is the dimensionality of the parameter space.  

As the average of the likelihood over the prior, the evidence is larger for a model if more of its parameter space is likely and smaller for a model where large areas of its parameter space have low likelihood values, even if the likelihood function is very highly peaked.  Thus, the evidence automatically implements Occam's razor: a simpler theory with a compact parameter space will have a larger evidence than a more complicated one, unless the latter is significantly better at explaining the data.  

The question of model selection between two models $H_0$ and $H_1$ can then be decided by comparing their respective posterior probabilities, given the observed data set $D$, via the model selection ratio $R$:

\begin{equation}
R= \frac{P(H_1\mid D)}{P(H_0\mid D)} = \frac{P(D \mid H_1)P(H_1)}{P(D\mid H_0)P(H_0)} = \frac{Z_1}{Z_0}\frac{P(H_1)}{P(H_0)},
\label{Eq:Rval}
\end{equation}
where $P(H_1)/P(H_0)$ is the a priori probability ratio for the two models, which can often be set to unity but occasionally requires further consideration.

\subsection{Nested Sampling and evaluating the evidence}

Evaluation of the multidimensional integral in Eq. \ref{eq:Evidence} is a challenging numerical task.  Standard techniques like thermodynamic integration \citep{Thermo} are extremely computationally expensive, which makes evidence evaluation  at least an order-of-magnitude more costly than parameter estimation.

The nested sampling approach \citep{2004AIPC..735..395S}  is a Monte-Carlo method targeted at the efficient calculation of the evidence, but also produces posterior inferences as a by-product.  In \cite{2009MNRAS.398.1601F} and \cite{2008MNRAS.384..449F} this nested sampling framework was built upon with the introduction of the MULTINEST algorithm, which provides an efficient means of sampling from posteriors that may contain multiple modes and/or large (curving) degeneracies, and also calculates the evidence.  Since its release MULTINEST has been used successfully in a wide range of astrophysical problems, from detecting the Sunyaev-Zel'dovich effect in galaxy clusters \citep{2012arXiv1210.7771C}, to inferring the properties of a potential stochastic gravitational wave background in pulsar timing array data \citep{2013PhRvD..87j4021L} (henceforth L13). This technique has greatly reduced the computational cost of Bayesian parameter estimation and model selection, and is employed in this paper.

\section{Shapelets}
\label{section:Shapelets}

A thorough description of the Shapelet formalism can be found in \cite{2003MNRAS.338...35R}, with astronomical uses being described in e.g, \cite{2004AJ....127..625K,2013MNRAS.430.2454L,2003MNRAS.338...48R}.  Here we give only an outline to aid later discussion.

Shapelets are described by a set of dimensionless basis functions, which in one dimension can be written:

\begin{equation}
\phi_n(x) \equiv \left[2^n\sqrt{\pi}n!\;\right]^{-1/2} H_n(x)\;\mathrm{e}^{-x^2/2},
\end{equation}
where $n$ is a non-negative integer, and $H_n$ is the Hermite polynomial of order $n$.  Therefore the lowest order shapelet is given by a standard Gaussian ($H_0(x) = 1$), with higher order terms represented by a Gaussian multiplied by the relevant polynomial.

These are then modified by a scale factor $\beta$ which is a free parameter to be fit for, in order to construct the dimensional basis functions:

\begin{equation}
B_n(x;\beta) \equiv \beta^{-1/2} \phi_n(\beta^{-1}x).
\end{equation}
These basis functions are orthonormal in that we can write

\begin{equation}
\int_{-\infty}^\infty \; \mathrm{d}x \; B_i(x;\beta)B_j(x;\beta) = \delta_{ij},
\end{equation}
where $\delta_{ij}$ is the Kronecker delta so that we can represent a function $f(x)$ as the sum:

\begin{equation}
\label{Eq:shapefunction}
f(x, \mathbf{c}, \beta) = \sum_{i\mathrm{=0}}^{n_{\mathrm{max}}} c_iB_i(x;\beta),
\end{equation}
where $c_i$ are shapelet coefficients, and $n_{\mathrm{max}}$ the number of shapelet basis vectors included in the model.  

Finally, the total integrated flux in the model profile, $F_{\mathrm{tot}}$, is given by

\begin{equation}
F_{\mathrm{tot}} =\;\int_{-\infty}^{\infty} \;\mathrm{d}x \; f(x) \; = \; \sum_{n\mathrm{=even}} \; c_n\big[2^{1-n}\pi^{\frac{1}{2}}\beta\big]^{\frac{1}{2}} {n \choose n/2}^{\frac{1}{2}},
\end{equation}
and the scale of the profile, $\sigma_{\mathrm{p}}$ is given by

\begin{eqnarray}
\sigma^2_{\mathrm{p}} &=& \;\int_{-\infty}^{\infty} \;\mathrm{d}x \; x^2f(x) \nonumber \\
                                       &=& \pi^{\frac{1}{4}}\beta^{\frac{5}{2}} F_{\mathrm{tot}}^{-1}\sum_{n\mathrm{=even}} \; 2^{\frac{1}{2}(1-n)}c_n\;(1+2n)\;{n \choose n/2}^{\frac{1}{2}}.
\end{eqnarray}

We use this definition of the scale to define the full width at half maximum (FWHM) of a particular profile shape as 2.3548$\sigma_{\mathrm{p}}$.

\section{Pulsar Timing Likelihood}
\label{Section:Like}

\begin{figure*}
\begin{center}$
\begin{array}{ccc}
\hspace{-0.7cm}
\includegraphics[width=60mm]{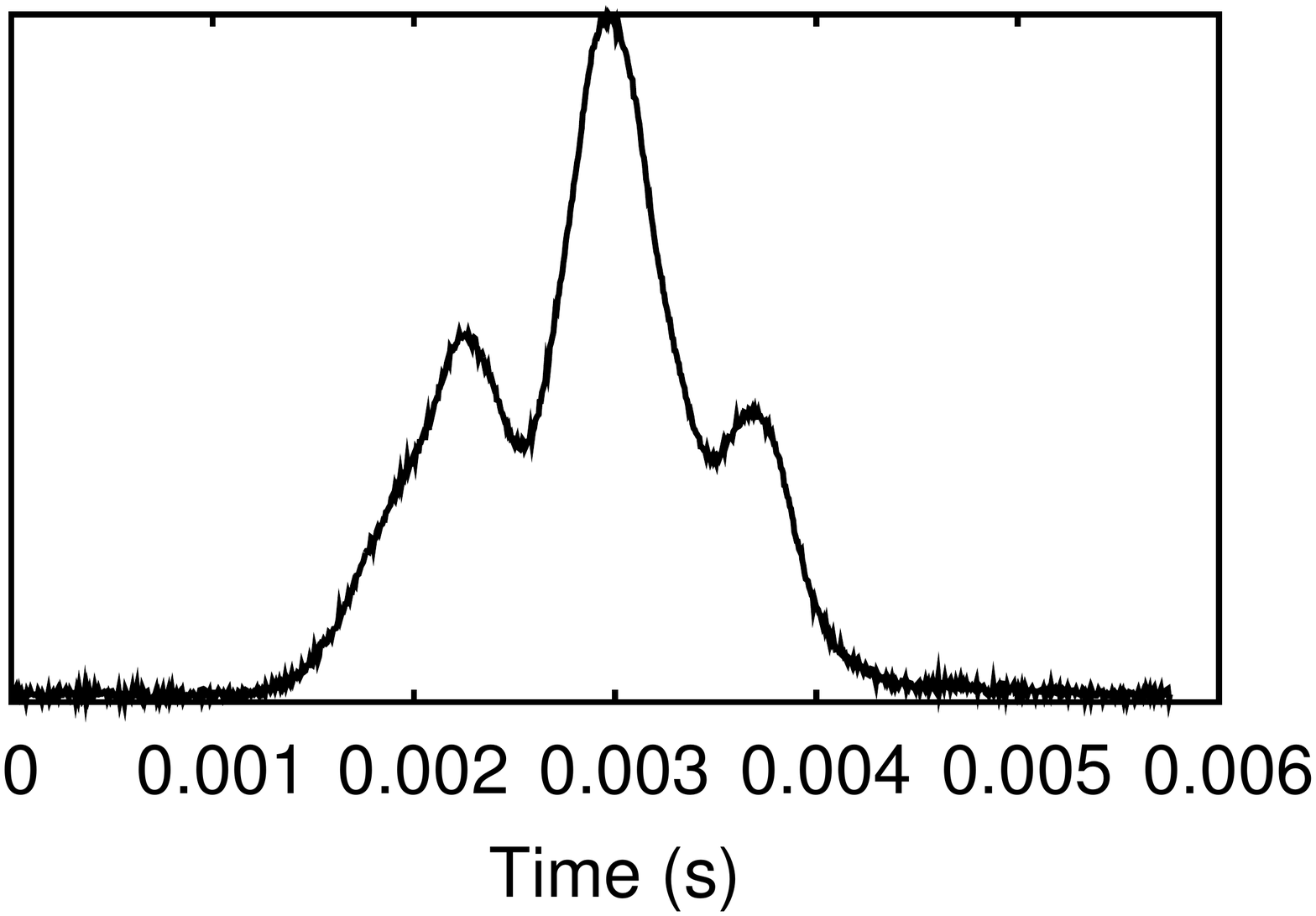} &
\includegraphics[width=60mm]{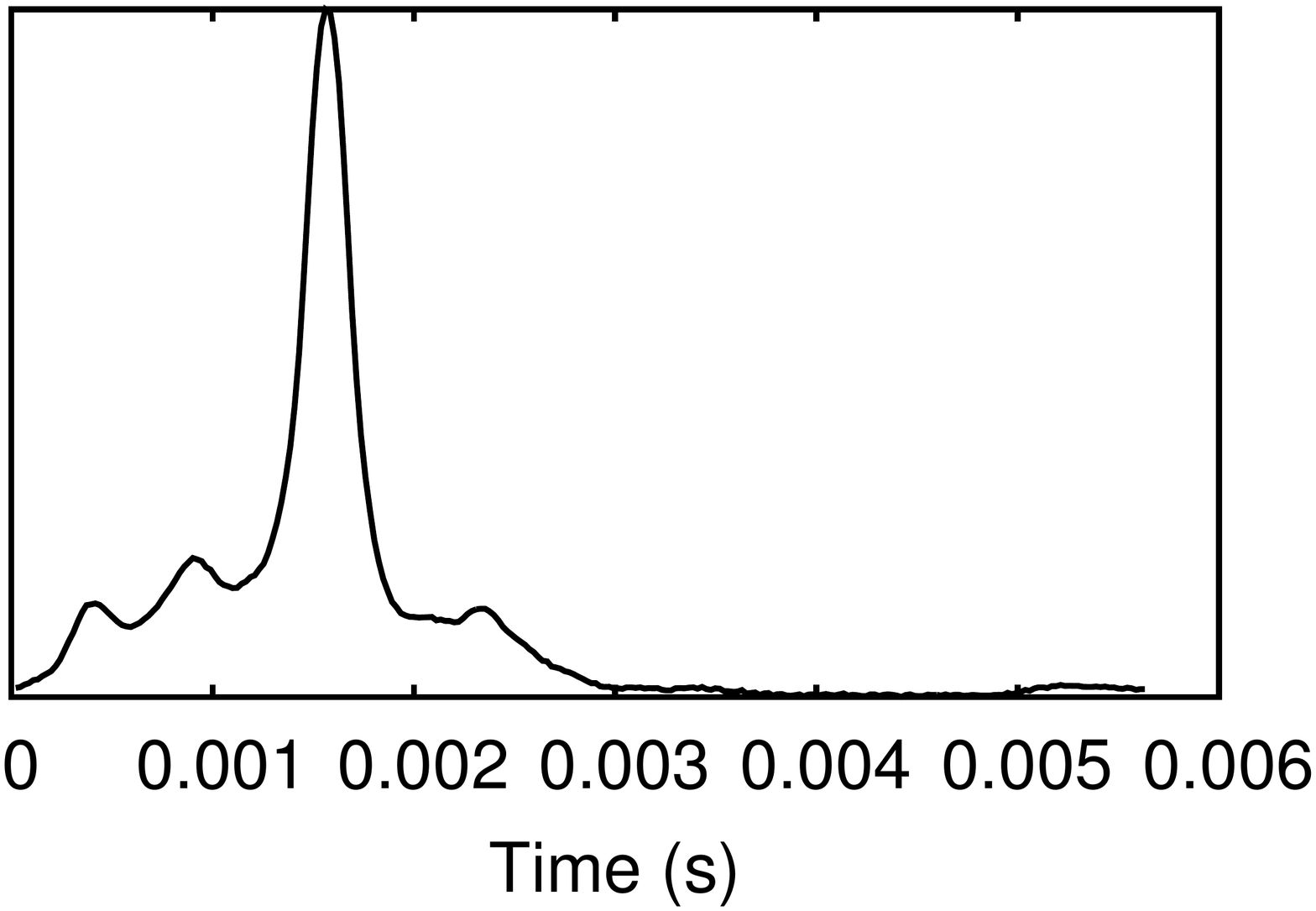} &
\includegraphics[width=60mm]{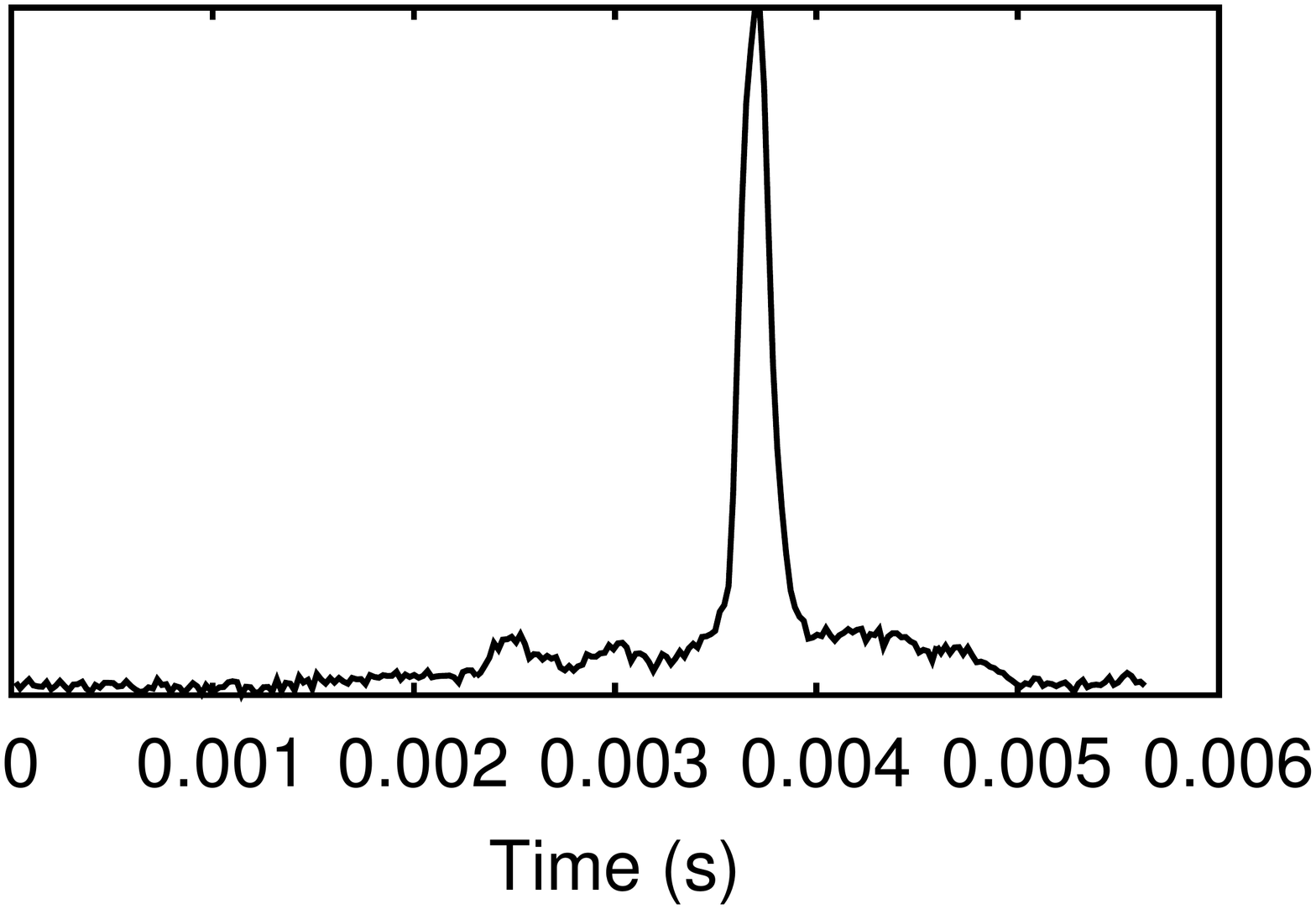} \\
\end{array}$
\end{center}
\caption[]{Integrated pulse profiles for PSR J0437-4715 at (left) 436MHz, (middle) 1440MHz, and (right) 4600MHz.  Profile data obtained from the ATNF pulsar catalogue \citep{2005AJ....129.1993M} \footnotemark.}
\label{figure:realprofiles}
\end{figure*}

We begin by considering our data, $\mathbf{d}$, a series of $n$ integrated pulse profiles that each consists of a set of $n_{\mathrm{bin}}$ values representing the flux density of the profile as measured at a set of times $\bmath{t}$.  Examples are these pulse profiles are shown in Fig. \ref{figure:realprofiles}. We represent $\mathbf{d}$ as a sum of both a deterministic and a stochastic component:

\begin{equation}
\mathbf{d}_{\mathrm{tot}} = \mathbf{d}_{\mathrm{det}} + \mathbf{d}_{\mathrm{sto}},
\end{equation}
where $\mathbf{d}_{\mathrm{tot}}$ represents the $n\times n_{\mathrm{bin}}$ total data points for the $n$ pulse profiles for a single pulsar, with $\mathbf{d}_{\mathrm{det}}$ and $\mathbf{d}_{\mathrm{sto}}$ the deterministic and stochastic contributions to the total respectively, where any contributions to the latter will be modelled as random Gaussian processes.
If we first consider the stochastic component of the signal, $\mathbf{d}_{\mathrm{sto}}$, to be described solely by an uncorrelated random Gaussian field with RMS $\sigma$, then the deterministic component, $\mathbf{d}_{\mathrm{det}}$,  will consist only of our shapelet model for the pulse profiles with the centroid position for each model profile determined using the set of arrival times $\bmath{\tau}(\bmath{\epsilon})$  predicted by the pulsar's timing model parameters, $\bmath{\epsilon}$.  With the inclusion of the timing model, we can rewrite Eq. \ref{Eq:shapefunction} to describe the shapelet model for a particular pulse $p$:

\begin{equation}
\label{Eq:newshapefunction}
s_p(t, \mathbf{c}, \beta, \epsilon) = \sum_{i\mathrm{=0}}^{n_{\mathrm{max}}} c_iB_i(t-\tau_p(\epsilon);\beta),
\end{equation}
with $\tau_p$ the arrival time for pulse $p$ predicted by the set of timing model parameters $\epsilon$.  We can then write the likelihood that the data is described only by the shapelet parameters, $\mathbf{\theta} \equiv (\mathbf{c}, \beta)$,  and the timing model parameters $\bmath{\epsilon}$ as:

\begin{equation}
\label{Eq:TimeLike}
\mathrm{Pr}(\mathbf{d} |\mathbf{\theta}, \bmath{\epsilon}) \propto \frac{1}{\sqrt{\mathrm{det}\mathbf{N}}} \exp{\left[-\frac{1}{2}(\mathbf{d} - \mathbf{s}(\mathbf{\theta}, \bmath{\epsilon}))^T\mathbf{N^{-1}}(\mathbf{d} - \mathbf{s}(\mathbf{\theta},\bmath{\epsilon}))\right]},
\end{equation}
where $\mathbf{s}(\mathbf{\theta}, \bmath{\epsilon})$ is the concatenated vector of all model pulse profiles and $\mathbf{N}$ is the diagonal white noise covariance matrix with elements $\sigma$.  To aid later discussion we note that (i) while in Eq. \ref{Eq:TimeLike} the amplitude of the noise in the profile data is taken to be a known quantity, additional parameters can be included into the analysis that scale or add in quadrature to $\sigma$, and (ii) with the shapelet parameters $\mathbf{\theta}$ one can then either define an arbitrary number of different profiles across time and frequency, or evolve a single model profile using some functional form (e.g. a scaling of $\beta$ with frequency) in order to encapsulate the evolution of the profile across the span of the data.

\subsection{Marginalising analytically over the shapelet amplitudes}
\label{section:Margins}

One can imagine two possible scenarios when fitting for the pulse profile using shapelets as described so far.  Either a single model profile with an average total amplitude  can be fit to the data, or a single profile can be fit, but with overall amplitude free to vary from one data profile to another.  In the latter case this will introduces $n$ additional free parameters (one per pulse profile in the dataset), however it is possible to marginalise analytically over these amplitude parameters without introducing large matrix operations and thus significantly decrease the size of parameter space we must sample over.  We first write the shapelet model vector $\mathbf{s}$ as the product of a set of $n$ amplitude parameters $\mathbf{A}$ and the $(n\times n_{\mathrm{bin}})\times n$ block diagonal matrix containing the $n$ repetitions of the normalised model profile $\mathbf{M}(\theta, \bmath{\epsilon})$, so that:

\begin{equation}
\mathbf{s} = \mathbf{MA}.
\end{equation}
In order to perform the marginalisation over the amplitudes $\mathbf{A}$, we first write the log of the likelihood in Eq \ref{Eq:TimeLike}, which, leaving out the determinant and denoting  $\mathbf{M}^T\mathbf{N}^{-1}\mathbf{M}$ as $\mathbf{\Sigma}$ and $\mathbf{M}^T\mathbf{N}^{-1}\mathbf{d}$ as $\mathbf{\hat{d}}$ is given by:
\begin{equation}
\label{Eq:LogL}
\log \mathrm{L} = -\frac{1}{2} \mathbf{d}^T\mathbf{N}^{-1} \mathbf{d} - \frac{1}{2}\mathbf{A}^T\mathbf{\Sigma}\mathbf{A} + \mathbf{\hat{d}}^T\mathbf{A}.
\end{equation}
Taking the derivative of $\log \mathrm{L}$ with respect to $\mathbf{A}$ gives us a vector of gradients:
\begin{equation}
\label{Eq:Grada}
\frac{\partial \log \mathrm{L}}{\partial \mathbf{A}} =  -\mathbf{A^T}\mathbf{\Sigma} + \mathbf{\hat{d}^T},
\end{equation}
which setting Eq \ref{Eq:Grada} equal to zero can be solved to give us the maximum likelihood vector of coefficients $\hat{\mathbf{A}}$:
\begin{equation}
\label{Eq:amax}
\hat{\mathbf{A}} = \mathbf{\Sigma}^{-1}\mathbf{\hat{d}}.
\end{equation}
Re-expressing Eq. \ref{Eq:LogL} in terms of $\hat{\mathbf{A}}$:

\begin{equation}
\label{Eq:beforemargin}
\log \mathrm{L} = -\frac{1}{2} \mathbf{d}^T\mathbf{N}^{-1} \mathbf{d} + \frac{1}{2}\hat{\mathbf{A}}^T\mathbf{\Sigma}\hat{\mathbf{A}} 
 -  \frac{1}{2}(\mathbf{A} - \hat{\mathbf{A}})^T\mathbf{\Sigma}(\mathbf{A} - \hat{\mathbf{A}}),
\end{equation}
we can integrate Eq. \ref{Eq:beforemargin} with respect to the $n$ elements in $\mathbf{A}$ to get the likelihood that the data is described by the remaining parameters alone, i.e:

\begin{equation}
\mathrm{Pr}(\mathbf{d} |\bmath{\theta}, \bmath{\epsilon}) = \int_{-\infty}^{+\infty} \mathrm{Pr}(\mathbf{d} |\bmath{\theta}, \bmath{\epsilon}, \bmath{A}) \mathrm{Pr}(\bmath{A}) \mathrm{d}\mathbf{A}.
\end{equation}
As only the third term in Eq. \ref{Eq:beforemargin} has a dependency on $\bmath{A}$, choosing a uniform prior on the amplitudes so that $\mathrm{Pr}(\bmath{A}) = 1$ this integral becomes:
\begin{eqnarray}
I &=& \int_{-\infty}^{+\infty}\mathrm{d}\mathbf{A}\exp\left[-\frac{1}{2}(\mathbf{A} - \hat{\mathbf{A}})^T\mathbf{\Sigma}(\mathbf{A} - \hat{\mathbf{A}})\right] \nonumber \\
&=& (2\pi)^n~\mathrm{det} ~ \mathbf{\Sigma}^{-\frac{1}{2}}.
\end{eqnarray}
Our marginalised probability distribution for parameters $\mathbf{\theta}$ and $\bmath{\epsilon}$ is then given by:

\begin{eqnarray}
\label{Eq:MarginAmp}
\mathrm{Pr}(\bmath{\epsilon}, \bmath{\theta} | \mathbf{d}) &\propto& \mathrm{Pr}(\mathbf{d} |\bmath{\theta}, \bmath{\epsilon})\mathrm{Pr}(\bmath{\theta}, \bmath{\epsilon})  \nonumber \\ 
&\propto& \frac{\mathrm{det} \left(\mathbf{\Sigma}\right)^{-\frac{1}{2}}}{\sqrt{\mathrm{det}\left(\mathbf{N}\right)}} 
\exp\left[-\frac{1}{2}\left(\mathbf{d}^T\mathbf{N}^{-1} \mathbf{d} - \mathbf{\hat{d}}^T\mathbf{\Sigma}^{-1}\mathbf{\hat{d}}\right)\right]. 
\end{eqnarray}

\section{Including additional stochastic parameters} 
\label{Section:Stochastic}

\subsection{Additional white noise}
\label{Section:White1}
\footnotetext{http://www.atnf.csiro.au/research/pulsar/psrcat}
In typical pulsar timing analysis the white noise is considered the sum of multiple terms, the radiometer noise associated with a given pulse profile as discussed in section \ref{Section:Like}, and additional terms  that represent sources of time independent noise.  These can include, for example, contributions from the high frequency tail of the pulsar's red spin noise power spectrum, or, jitter noise that results from the time averaging of a finite number of single pulses to form each TOA (see e.g. \cite{2014MNRAS.443.1463S}).  For simplicity we will refer to these quadrature terms simply as `EQUAD'.

One approach to including these EQUAD parameters in the likelihood is to use a set of $n$ free parameters $j_i$ that each represent a shift in the arrival time of a given pulse profile $i$, and then having a prior on those parameters that describes their underlying distribution.

If we first assume a Gaussian distribution on the parameters $\mathbf{j}$ we can write our likelihood as:

\begin{eqnarray}
\label{Eq:Pulsarjitterlike}
\mathrm{Pr}(\mathbf{d} | \mathbf{\theta}, \bmath{\epsilon}, \mathbf{j},  \mathbf{J}) &=& \mathrm{Pr}(\mathbf{d} |\mathbf{\theta}, \bmath{\epsilon}, \mathbf{j}) \times \mathrm{Pr}(\mathbf{j} | \mathbf{J})
\end{eqnarray}
where $\mathrm{Pr}(\mathbf{d} |\mathbf{\theta}, \bmath{\epsilon}, \mathbf{j})$ has the same functional form as in Eq. \ref{Eq:TimeLike}, but with the arrival time of profile $i$, $\tau_i$, now a function of the shift parameter $j_i$, and 

\begin{equation}
\mathrm{Pr}(\mathbf{j} | \mathbf{J}) = \frac{1}{\sqrt{(2\pi)^{n}\mathrm{det}~\mathbf{J}}}\exp\left[-\frac{1}{2}\mathbf{j}^T\mathbf{J^{-1}}\mathbf{j}\right],
\end{equation}
with $\mathbf{J}$ a diagonal matrix, where each element is equal to $\gamma^2$, the variance of the shift parameters $\mathbf{j}$, with $\gamma$ a free parameter to be fit for.  We note that, as in \cite{2014arXiv1405.2460L}, this formalism allows for the parameterisation of a non-Gaussian distribution in the parameters $\mathbf{j}$, however we do not pursue this option further here.

\subsection{Additional red spin noise}

In order to include additional red noise processes we begin by taking the same approach as that given in \cite{2013arXiv1310.2120L} (henceforth L13), which we will describe in brief below to aid further discussion.  Writing the red noise component of the stochastic signal, which we will denote $\mathbf{d}_{\mathrm{red}}$, in terms of its Fourier coefficients $\mathbf{a}$ so that $\mathbf{d}_{\mathrm{red}} = \mathbf{F}\mathbf{a}$ where $\mathbf{F}$ denotes the Fourier transform such that for frequency $\nu$ and time $t$ we will have both:

\begin{equation}
\label{Eq:FMatrix}
F(\nu, \tau_i) = \frac{1}{T}\sin\left(2\pi\nu \tau_i\right),
\end{equation}
and an equivalent cosine term.  Here $T$ represents the total observing span for the pulsar, $\nu$ the frequency of the signal to be sampled and $\tau_i$ the model TOA for profile $i$.  Defining the number of coefficients to be sampled by $n_{\mathrm{max}}$, we can then include the set of frequencies with values $n/T$, where $n$ extends from 1 to $n_{\mathrm{max}}$.
For typical pulsar timing array (PTA) data \cite{2012MNRAS.423.2642L} show that a low frequency cut off of $1/T$ is sufficient to accurately describe the expected long term variations present in the data. If necessary though it is also possible to specify arbitrary sets of frequencies such that terms with $\nu << 1/T$ can be included in the model, or to allow noise terms where the frequency itself is a free parameter. 

For a single pulsar the covariance matrix $\bmath{\varphi}$ of the Fourier coefficients $\mathbf{a}$ will be diagonal, with components

\begin{equation}
\label{Eq:BPrior}
\varphi_{ij} = \left< a_ia_j^*\right> = \varphi_{i}\delta_{ij},
\end{equation}
where there is no sum over $i$, and the brackets $\left<..\right>$ equate to the expectation value such that the set of coefficients $\{\varphi_{i}\}$ represent the theoretical power spectrum of the red noise signal present in the timing data.

As discussed in L13, whilst Eq \ref{Eq:BPrior} states that the Fourier modes are orthogonal to one another, this does not mean that we assume they are orthogonal in the time domain where they are sampled, and it can be shown that this non-orthogonality is accounted for within the likelihood.  Instead, in Bayesian terms, Eq. \ref{Eq:BPrior} represents our prior knowledge of the power spectrum coefficients within the data.  We are therefore stating that, whilst we do not know the form the power spectrum will take, we know that the underlying Fourier modes are still orthogonal by definition, regardless of how they are sampled in the time domain.  It is here then that, should one wish to fit a specific model to the power spectrum coefficients at the point of sampling, such as a broken, or single power law, the set of coefficients $\{\varphi_{i}\}$ should be given by some function $f(\Theta)$, where we sample from the parameters $\Theta$ from which the power spectrum coefficients $\{\varphi_{i}\}$ can then be derived.

We can then then use the signal realisation of the red noise process given by $\mathbf{Fa}$ to alter the model TOAs given by the timing model,  $\tau(\bmath{\epsilon})$ such that:

\begin{equation}
\bmath{\hat{\tau}}(\bmath{\epsilon}, \bmath{a}) =  \bmath{\tau}(\bmath{\epsilon}) - \mathbf{Fa},
\end{equation}
enabling us to write the joint probability density of the timing model, power spectrum coefficients and the signal realisation, Pr$(\bmath{\epsilon}, \bmath{\theta}, \bmath{\varphi}, \mathbf{a} \;|\; \mathbf{d})$, as:

\begin{eqnarray}
\label{Eq:Prob}
\mathrm{Pr}(\bmath{\epsilon}, \bmath{\theta},\bmath{\varphi}, \mathbf{a} \;|\; \mathbf{d}) \; &\propto& \; \mathrm{Pr}(\mathbf{d} |  \bmath{\epsilon}, \bmath{\theta}, \mathbf{a}) \; \\\ \nonumber
&\times & \mathrm{Pr}(\mathbf{a} | \bmath{\varphi}) \; \mathrm{Pr}(\bmath{\varphi}). \nonumber
\end{eqnarray}
For our choice of $\mathrm{Pr}(\bmath{\varphi})$ we use an uninformative prior that is uniform in $\log_{10}$ space, and draw our samples from the parameter $\rho_i = \log_{10}(\varphi_i)$ instead of $\varphi_i$.  Given this choice of prior the conditional distributions that make up Eq. \ref{Eq:Prob} can be written:

\begin{eqnarray}
\label{Eq:ProbTime}
& &\mathrm{Pr}(\mathbf{d} |\bmath{\epsilon}, \bmath{\theta}, \mathbf{a}) \; \propto \; \frac{1}{\sqrt{\mathrm{det}(\mathbf{N})}}  \\
& \times & \exp\left[-\frac{1}{2}(\mathbf{d} -  \bmath{s}(\bmath{\theta}, \bmath{\epsilon}, \bmath{a}))^T\mathbf{N}^{-1}(\mathbf{d} -  \bmath{s}(\bmath{\theta}, \bmath{\epsilon}, \bmath{a}))\right] \nonumber
\end{eqnarray}
and:

\begin{equation}
\label{Eq:ProbFreq}
\mathrm{Pr}(\mathbf{a}\; | \;\bmath{\rho}) \; \propto \; \frac{1}{\sqrt{\mathrm{det}\bmath{\varphi}}} \exp\left[-\frac{1}{2}\mathbf{a}^{T}\bmath{\varphi}^{-1}\mathbf{a}\right].
\end{equation}

\subsection{Including dispersion measure variations}

The plasma located in the interstellar medium (ISM), as well as in solar winds and the ionosphere can result in delays in the propagation of the pulse signal between the pulsar and the observatory, an effect that appears as a red noise signal in the timing residuals.  

Unlike other red noise signals however, the severity of the observed dispersion measure variations is dependent upon the observing frequency, and as such we can use this additional information to isolate the component of the red noise that results from this effect.

In particular, the group delay $t_g(f)$ for a frequency $f$ is given by the relation:

\begin{equation}
t_g(f) = DM/(Kf^2),
\end{equation}
where the dispersion constant $K$ is given by:

\begin{equation}
K \equiv 2.41 \times 10^{-16}~\mathrm{Hz^{-2}~cm^{-3}~pc~s^{-1}}
\end{equation}
and the dispersion measure is defined as the integral of the electron density $n_e$ from the Earth to the pulsar:

\begin{equation}
DM = \int_0^L n_e \mathrm{d}l.
\end{equation}

Dispersion measure corrections can be included in the analysis as an additional set of stochastic parameters in almost the same was as the frequency independent spin noise.  We begin by first defining a vector $\bmath{D}$ of length equal to the number of pulse profiles $n$ for a given pulsar as:

\begin{equation}
D_i = 1/(Kf^2_i)
\end{equation}
for observation $i$ with observing frequency $f_i$.

We then write the basis vectors that describe the dispersion measure Fourier modes as:

\begin{equation}
F^{DM}(\nu,\tau_i) = \frac{1}{T}\sin\left(2\pi\nu \tau_i\right)D_i
\end{equation}
and an equivalent cosine term, where $T$ is the length of the observing timespan, and $\nu$ denotes the frequency of the signal to be parameterised as before, where the set of frequencies to be included is defined in the same way as for the red spin noise.  Unlike when modelling the red spin noise, we no longer have the quadratic in the timing model to act as a proxy to the low frequency ($\nu < 1/T$) DM variations in our data.  As such these terms must be accounted for either by explicitly including these low frequencies in the model, or by including a quadratic in DM to act as a proxy, as with the red noise, defined as: 

\begin{equation}
 Q_{\mathrm{DM}}(\tau_i)= \alpha_0 \tau_iD_i + \alpha_1 \tau_i^2D_i
\end{equation}
with $\alpha_{0,1}$ free parameters to be fit for, and $\tau_i$ the barycentric arrival time for profile $i$. This is most simply done by including these terms in the set of timing model parameters $\bmath{\epsilon}$.  If these terms are not included in the model then power from frequencies lower than $1/T$ will be absorbed by the Fourier coefficients included in $\mathbf{F_{\mathrm{DM}}}$, biasing the estimated power spectrum.  

As with the red noise we can then include the DM signal realisation in our model TOAs:

\begin{equation}
\bmath{\hat{\tau}}(\bmath{\epsilon}, \bmath{a_{\mathrm{red}}}, \bmath{a_{\mathrm{DM}}}) =  \bmath{\tau}(\bmath{\epsilon}) - \mathbf{F_{\mathrm{red}}a_{\mathrm{red}}} - \mathbf{F_{\mathrm{DM}}a_{\mathrm{DM}}},
\end{equation}
where we have factored the quadratic $Q_{\mathrm{DM}}$ into the timing model $\bmath{\tau}(\bmath{\epsilon})$. Finally we then define the matrix of DM power spectrum coefficients $\bmath{\varphi_{\mathrm{DM}}}$ such that:

\begin{equation}
\label{Eq:ProbDMFreq}
\mathrm{Pr}(\mathbf{a_{\mathrm{DM}}}\; | \;\bmath{\rho_{\mathrm{DM}}}) \; \propto \; \frac{1}{\sqrt{\mathrm{det}\bmath{\varphi_{\mathrm{DM}}}}} \exp\left[-\frac{1}{2}\mathbf{a_{\mathrm{DM}}}^{T}\bmath{\varphi_{\mathrm{DM}}}^{-1}\mathbf{a_{\mathrm{DM}}}\right].
\end{equation}
We note here that, as with the red noise, if desired additional terms can be added into the DM Fourier matrix $\mathbf{F_{\mathrm{DM}}}$ to model, for example, additional annual variations in the data.

\section{Extending the profile model}

The simplicity of our method as outlined in Sections \ref{Section:Like} and \ref{Section:Stochastic}, allows for many possible extensions to include greater complexity in the pulse profile model.  In this section we will briefly consider 3 such extensions, i) Including correlated noise in the folded profile data, ii) Including pulse broadening effects due to scattering in the interstellar medium, and finally iii) Profile evolution.  We will incorporate the third of these in Simulation 3 in Section \ref{Section:Sim4}.

\subsection{Including correlated noise in the folded data}
\label{Section:RedFold}

Thus far we have assumed that the noise in the integrated pulse profiles that make up our data vector $\bmath{d}$ is described entirely by an uncorrelated random Gaussian field. In actual data however this might not be the case, with interference, or other effects resulting a power spectrum for the noise that is not white.  We can include this in our analysis by defining an $F$-matrix as in Eq. \ref{Eq:FMatrix} using the time stamps for $n_{\mathrm{bin}}$ data points that make up the pulse profile.  Eq. \ref{Eq:TimeLike} can then be extended simply as:

\begin{eqnarray}
\label{Eq:ProbTime2}
& &\mathrm{Pr}(\mathbf{d} |\bmath{\epsilon}, \bmath{\theta}, \mathbf{a_p}) \; \propto \; \frac{1}{\sqrt{\mathrm{det}(\mathbf{N})}}  \\
& \times & \exp\left[-\frac{1}{2}(\mathbf{d} -  \bmath{s}(\bmath{\theta}, \bmath{\epsilon}) - \bmath{F_pa_p})^T\mathbf{N}^{-1}(\mathbf{d} -  \bmath{s}(\bmath{\theta}, \bmath{\epsilon}) - \bmath{F_pa_p})\right] \nonumber
\end{eqnarray}
where $\bmath{F_pa_p}$ are the Fourier matrices and coefficients describing the noise in the profile, and:

\begin{equation}
\label{Eq:ProbFreq2}
\mathrm{Pr}(\mathbf{a_p} | \bmath{\rho_p}) \; \propto \; \frac{1}{\sqrt{\mathrm{det}\bmath{\varphi_p}}} \exp\left[-\frac{1}{2}\mathbf{a_p}^{T}\bmath{\varphi_p}^{-1}\mathbf{a_p}\right].
\end{equation}
where $\bmath{\varphi_p}$ describes the power spectrum coefficients of the profile noise, where as before we take a $\log$-uniform prior on the power spectrum coefficients to reflect our uncertainty in the values they will take, defining $\bmath{\rho_p} = \log_{10}[\bmath{\varphi_p}]$.

Because each profile is independent, adding in a set of Fourier coefficients for every profile, even if the underlying power spectrum is modelled to be the same, can lead to a prohibitive increase in the dimensionality of the problem.  As such, one can marginalise analytically over the coefficients $\mathbf{a_p}$ analytically and simply fit for the power spectrum coefficients present in the matrix $\bmath{\varphi_p}$.

In order to perform the marginalisation over the Fourier coefficients $\mathbf{a_p}$, we follow a similar process as described in Section \ref{section:Margins}.  Now, however, we denote $\mathbf{d} -  \bmath{s}(\bmath{\theta}, \bmath{\epsilon}, \bmath{a})$ as  $\mathbf{\delta d}$,  $(\mathbf{F_p}^T\mathbf{N}^{-1}\mathbf{F_p} + \mathbf{\varphi_p}^{-1})$ as $\mathbf{\bar{\Sigma}}$ and $\mathbf{F_p}^T\mathbf{N}^{-1}\mathbf{\delta d}$ as $\mathbf{\bar{d}}$ and finish with the expression:
\begin{eqnarray}
\label{Eq:Margin}
\mathrm{Pr}(\bmath{\epsilon}, \bmath{\theta}, \bmath{\rho_p} | \mathbf{d}) &\propto& \frac{\mathrm{det} \left(\mathbf{\bar{\Sigma}}\right)^{-\frac{1}{2}}}{\sqrt{\mathrm{det} \left(\mathbf{\varphi_p}\right)~\mathrm{det}\left(\mathbf{N}\right)}} \nonumber \\
&\times&\exp\left[-\frac{1}{2}\left(\mathbf{\delta d}^T\mathbf{N}^{-1} \mathbf{\delta d} - \mathbf{\bar{d}}^T\mathbf{\bar{\Sigma}}^{-1}\mathbf{\bar{d}}\right)\right]. 
\end{eqnarray}

\subsection{Including pulse broadening}
\label{Section:Broaden}

When a pulse propogates through a non-uniform, ionized medium it will be scattered.  The result of this scattering is that photons from a single pulse will follow different paths through the medium, leading to a broadening of the profile.  This broadening has a characteristed time scale $\tau_d$ that depends on the properties of the scattering medium (for details see e.g. \cite{1990ARA&A..28..561R}), and is highly dependent on the observing frequency, with a typical scaling relation of $\alpha_{\nu} \sim \nu^4$ \cite{2004ApJ...605..759B}.

We can include the effects of pulse broadening in our analysis by considering the pulse broadening function (PBF).  One simple example of a PBF if the scattering is considered to take place only in a thin screen is given by an exponential decay \citep{1970Sci...168.1453C}:

\begin{equation}
\mathrm{PBF}(t) \sim \exp(t/\tau_d).
\end{equation}
The observed pulse profile $\bmath{s}$ is then given by the convolution of this PBF, with $\tau_d$ and $\alpha_{\nu}$ additional free parameters in the analysis, and our shapelet model for the intrinsic pulse profile shape.

\subsection{Including pulse evolution}

Finally we consider evolution in the pulse profile, whether in frequency or in time, either of which can be included trivially in this framework.  This can be done either be specifying an arbitrary number of independent profiles that are evaluated during different observational epochs, or observing bands, or by defining some functional form that modifies a single pulse profile over the course of the dataset, such as a scaling of the pulse width $\beta$ with frequency.

In this way jumps are naturally incorporated into the analysis as a change in the centroid of the pulse profile used, although naturally additional jumps can still be included as a separate parameter, however they should be defined so as not to introduce covariance between the jump and profile parameters.

\section{Evaluating the likelihood}
\label{Section:Generate}

In the previous section we have defined our likelihood as a function of the timing model parameters $\bmath{\epsilon}$, shapelet parameters $\bmath{\theta}$ and additional noise parameters $\bmath{a},\bmath{\rho}$.  We will now describe more qualitatively the process by which we calculate the model TOAs required to evaluate the shapelet model for each likelihood calculation.  When performing traditional timing analysis, software packages such as Tempo2 or TempoNest take a parameter file and a tim file as input.  The former of these defines the timing model parameters to be used in the analysis, and the latter defines a set of site arrival times (SATs), that represent the measured data.

\begin{figure*}
\begin{center}$
\begin{array}{c}
\hspace{-0.7cm}
\includegraphics[width=100mm]{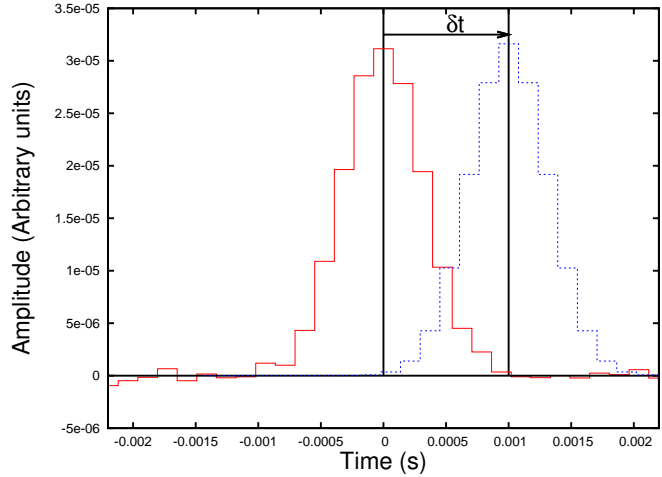} \\
\end{array}$
\end{center}
\caption{Graphical representation of the a single likelihood evaluation in GPTA for a given set of timing model parameters $\bmath{\epsilon}$, and profile parameters $\bmath{\theta}$. The solid red line shows the data, a pulse profile with additional white noise.  The vertical line at $x=0$ represents the SAT present in the tim file that corresponds to that pulse, transformed to the solar system barycenter.  Values on the $x$ axis are the separation from this BAT in seconds.  Note that the exact value of the SAT is not important, it is simply used as a flagpost in parameter space against which to define the model TOA used in GPTA.   A model TOA is generated by using a set of timing model parameters $\bmath{\epsilon}$ to obtain a residual $\delta t$ relative to the existing BAT.  The model profile (blue dashed line) can then be evaluated at this model TOA.  In this way the timing model can be used to generate model TOAs, and the model pulse profile can be evaluated concurrently.}
\label{figure:gentoa}
\end{figure*}

In GPTA we use these previously obtained SATs as `flagposts' in time against which we can define the model arrival times at which to evaluate the model pulse profiles.  Given a set of timing model parameters, we can transform these initial SATs to barycentric time of arrivals (BATs).  We then assume that the transformation for a given SAT holds for the entire pulse profile associated with that SAT.  The timing model then allows us to calculate a residual $\delta t$ for every BAT.  We use this residual to define a new model TOA, and evaluate the model pulse profile at that point, calculating the likelihood with respect to the folded data.  This process is shown figuratively in Fig. \ref{figure:gentoa}.

\section{Application to simulations}
\label{Section:simulations}

We now perform a series of 3 simulations designed to show the efficacy of the method described in Section \ref{Section:Like}.  We will describe each simulation in brief below and then discuss the results in more detail in Sections \ref{Section:Sim1} to \ref{Section:Sim4}.

\begin{table}
\caption{Simulated timing model parameter values for PSR J0030+0451}
\begin{tabular}{lc}
\hline\hline
Model Parameter & Injected value\\
\multicolumn{2}{c}{} \\ 
\hline
Right ascension, $\alpha$\dotfill &  00:30:27.4299630\\ 
Declination, $\delta$\dotfill & +04:51:39.75230 \\ 
Pulse frequency, $\nu$ (s$^{-1}$)\dotfill &  205.530696088273125 \\ 
First derivative of pulse frequency, $\dot{\nu}$ (s$^{-2}$)\dotfill & $-$4.3060388$\times 10^{-16}$ \\ 
Proper motion in right ascension, $\mu_{\alpha}$ (mas\,yr$^{-1}$)\dotfill & $-$4.054 \\ 
Proper motion in declination, $\mu_{\delta}$ (mas\,yr$^{-1}$)\dotfill & $-$5.034 \\ 
Parallax, $\pi$ (mas)\dotfill & 4.0229 \\ 
\hline
\end{tabular}
\label{Table:Simvals}
\end{table}

\begin{description}
  \item[{\bf Simulation 1:}] An 8 year dataset for the isolated pulsar PSR J0030+0451 using the parameters given in Table \ref{Table:Simvals}, with observations taken to occur at $\sim$ 2 week intervals. White noise is added to the profile data such that the TOAs have an average $1\sigma$ uncertainty of $3.5\times10^{-6}$s.\\
  
  \item[{\bf Simulation 2:}]  As Simulation 1, but for a 5 year timespan, with TOA uncertainties of $10^{-7}$s and with the addition of red timing noise to the dataset of the form $\mathrm{P}(f) = Af^{-4.33}$.\\

  \item[{\bf Simulation 3:}]  As Simulation 1, but we separate the data into two observing frequencies and evolve the pulse profile between frequencies.\\

\end{description}

\begin{table*}
\caption{Parameter estimates for the four simulations of PSR J0030+0451}
\begin{tabular}{lcc}
\hline\hline
\multicolumn{3}{c}{Simulation 1} \\ 
\hline
Model Parameter & GPTA & TempoNest \\
\hline
Right ascension, $\alpha$\dotfill &  00:30:27.4299(3) & 00:30:27.4298(3)\\ 
Declination, $\delta$\dotfill & +04:51:39.757(9) & +04:51:39.758(9)\\ 
Pulse frequency, $\nu$ (s$^{-1}$)\dotfill  & 205.530696088272(2) & 205.530696088272(2)\\ 
First derivative of pulse frequency, $\dot{\nu}$ (s$^{-2}$)\dotfill & $-$4.30603(18)$\times 10^{-16}$ & $-$4.30603(18)$\times 10^{-16}$\\ 
Proper motion in right ascension, $\mu_{\alpha}$ (mas\,yr$^{-1}$)\dotfill & $-$4.0(9) & $-$4.1(9)\\ 
Proper motion in declination, $\mu_{\delta}$ (mas\,yr$^{-1}$)\dotfill & $-$5(2) & $-$5(2)\\ 
Parallax, $\pi$ (mas)\dotfill & 3.3(5) & 3.4(5))\\ 
Profile $\sigma$ (s) \dotfill & 3.14(2)$\times 10^{-3}$ & - \\ 
\hline
\hline
\multicolumn{3}{c}{Simulation 2} \\ 
\hline
Model Parameter & GPTA & TempoNest \\
\hline
Right ascension, $\alpha$\dotfill &  00:30:27.42991(7) & 00:30:27.42992(7)\\ 
Declination, $\delta$\dotfill & +04:51:39.749(3) & +04:51:39.748(3)\\ 
Pulse frequency, $\nu$ (s$^{-1}$)\dotfill  & 205.530696088275(10) & 205.530696088277(11)\\ 
First derivative of pulse frequency, $\dot{\nu}$ (s$^{-2}$)\dotfill & $-$4.3040(14)$\times 10^{-16}$ & $-$4.3044(14)$\times 10^{-16}$\\ 
Proper motion in right ascension, $\mu_{\alpha}$ (mas\,yr$^{-1}$)\dotfill & $-$3.8(4) & $-$3.8(4)\\ 
Proper motion in declination, $\mu_{\delta}$ (mas\,yr$^{-1}$)\dotfill & $-$4.8(10) & $-$4.7(10)\\ 
Parallax, $\pi$ (mas)\dotfill & 4.05(6) & 4.07(6)\\ 
Log$_{10}$ Red noise amplitude \dotfill & $-$3.1(3) & $-$3.1(3) \\ 
Red noise spectral index\dotfill & 4.6(13) & 4.5(12)\\ 
\hline
\hline
\multicolumn{3}{c}{Simulation 3} \\ 
\hline
Model Parameter & GPTA with two profiles & GPTA with one profile \\
\hline
Right ascension, $\alpha$\dotfill &  00:30:27.4299639(9) & 00:30:27.4299640(9)\\ 
Declination, $\delta$\dotfill & +04:51:39.75227(3) & +04:51:39.75227(3)\\ 
Pulse frequency, $\nu$ (s$^{-1}$)\dotfill  & 205.530696088273128(8) &205.530696088273129(8) \\ 
First derivative of pulse frequency, $\dot{\nu}$ (s$^{-2}$)\dotfill & $-$4.3060387(6)$\times 10^{-16}$ & $-$4.3060388(6)$\times 10^{-16}$\\ 
Proper motion in right ascension, $\mu_{\alpha}$ (mas\,yr$^{-1}$)\dotfill & $-$4.052(3) & $-$4.052(3)\\ 
Proper motion in declination, $\mu_{\delta}$ (mas\,yr$^{-1}$)\dotfill & $-$5.038(6) & 5.036(5)\\ 
Parallax, $\pi$ (mas)\dotfill & 4.0231(16) & 4.0232(16)\\ 
\hline
\hline
\end{tabular}
\label{Table:Sim1}
\end{table*}

\subsection{Simulation 1}
\label{Section:Sim1}

We first consider the simplest possible case.  We simulate an 8 year dataset for the isolated pulsar PSR J0030+0451 using the parameters given in Table \ref{Table:Simvals}, with observations taken to occur at $\sim$ 2 week intervals.  We use a Gaussian pulse profile with $\sigma = 3.139\times 10^{-4}$ seconds for each TOA, with an amplitude of $3.16\times 10^{-5}$ in arbitrary units, each of which is sampled in 31 bins.  We then add Gaussian noise to each profile that is uncorrelated between bins with an rms such that the integrated signal--to--noise in each profile is $\sim 85$.  We form the TOAs by fitting the injected model Gaussian to each profile to obtain a maximum likelihood and associated $1\sigma$ uncertainty, which results in an average TOA error of $3.5\times10^{-6}$ seconds.  We then perform the analysis using GPTA using the profile data, and with TempoNest using the TOAs in order to compare the result.  In this instance we would expect both methods to be completely consistent as the TOA values and uncertainties estimated by fitting the Gaussian profile to the simulated folded data will be accounted for correctly in the TempoNest analysis.

\begin{figure*}
\begin{center}$
\begin{array}{ccc}
%
\includegraphics[width=50mm]{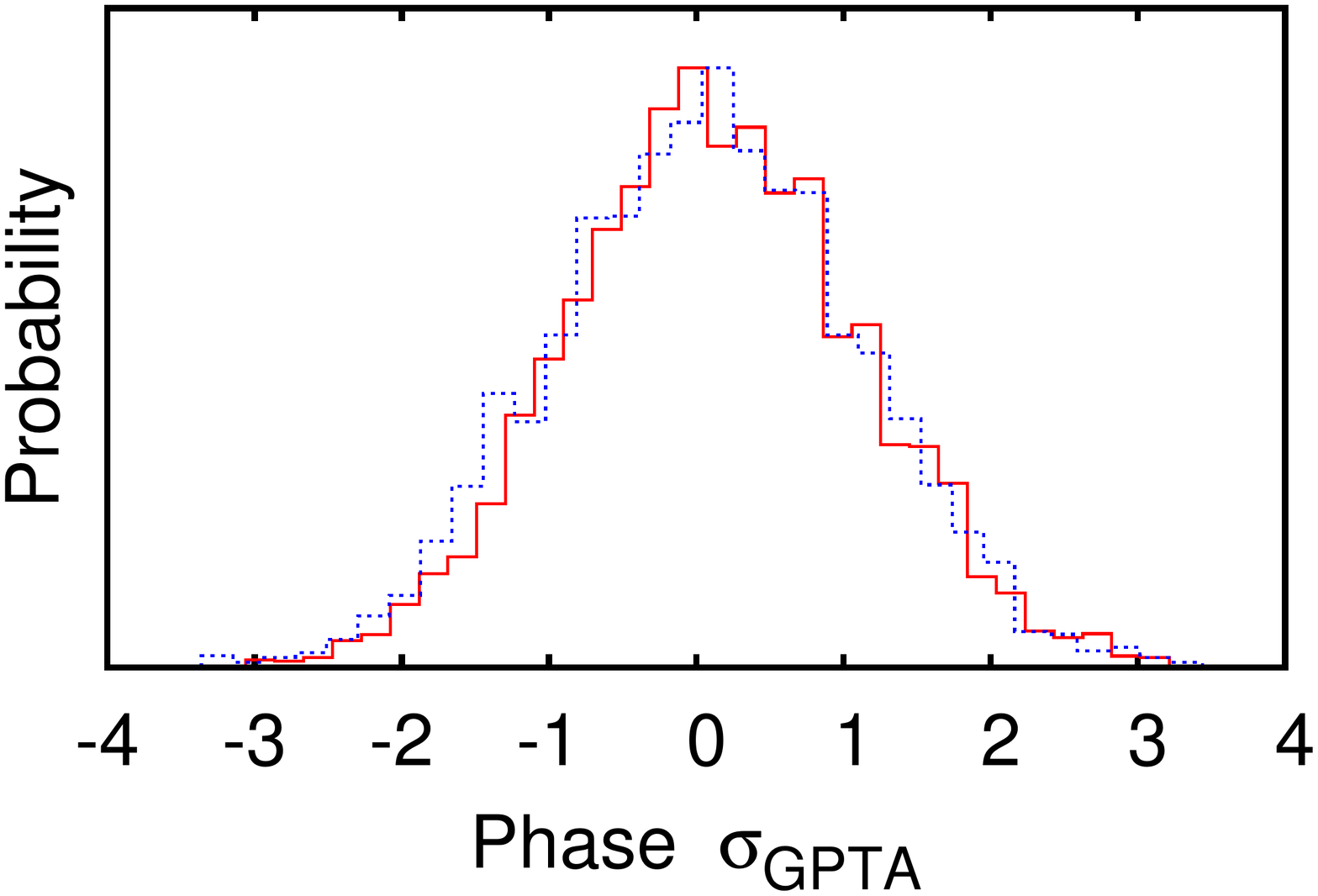} &
\includegraphics[width=50mm]{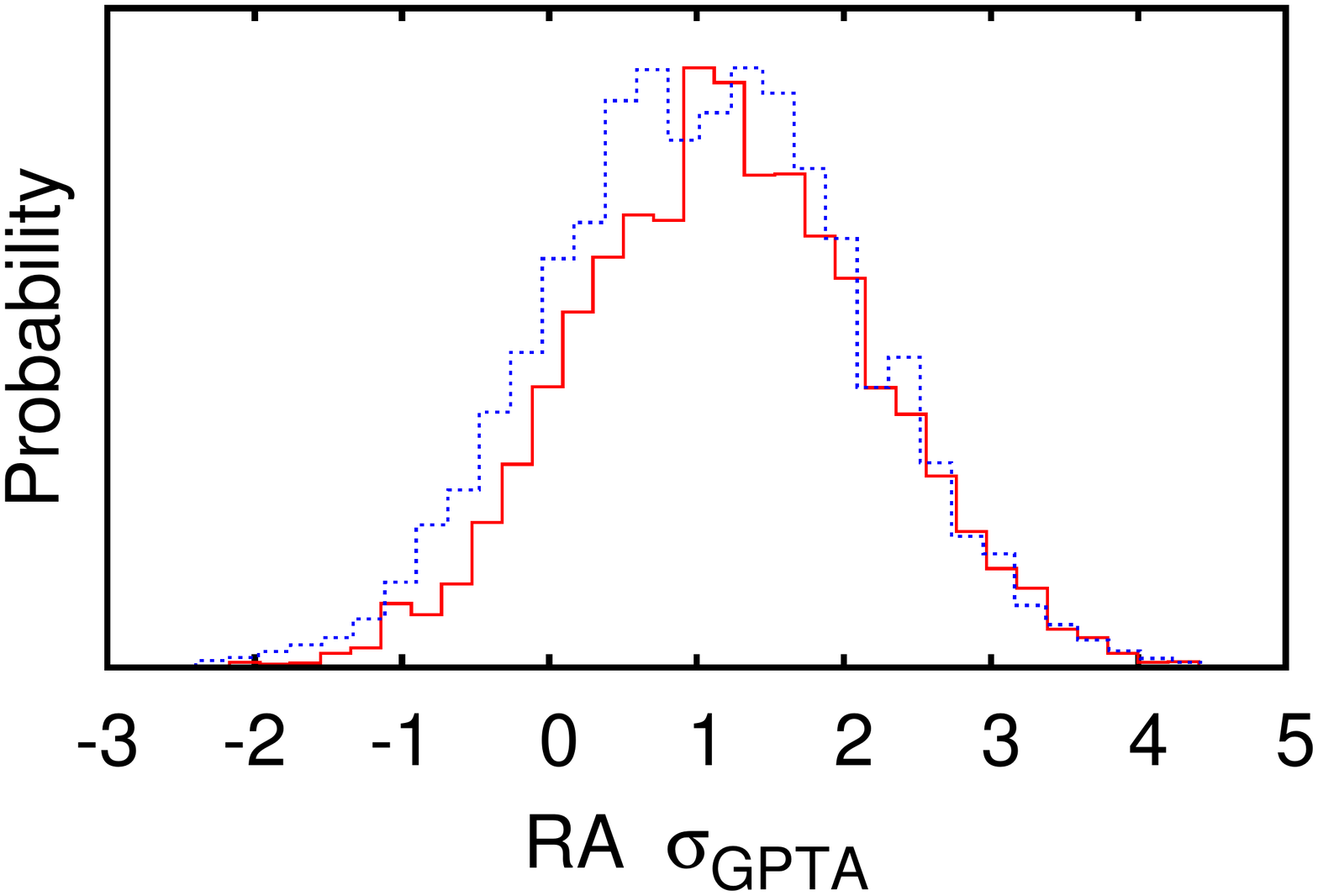} &
\includegraphics[width=50mm]{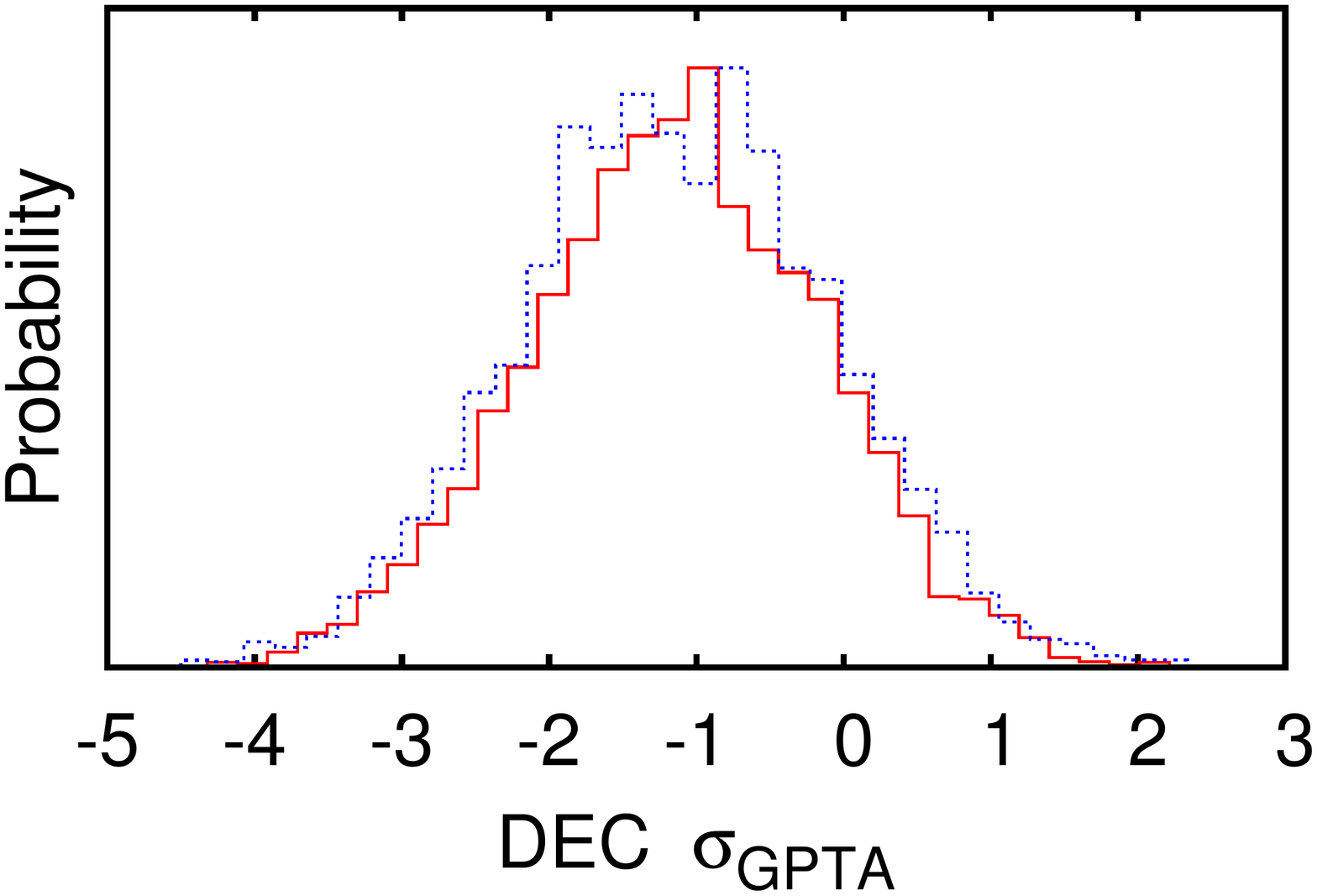} \\
\includegraphics[width=50mm]{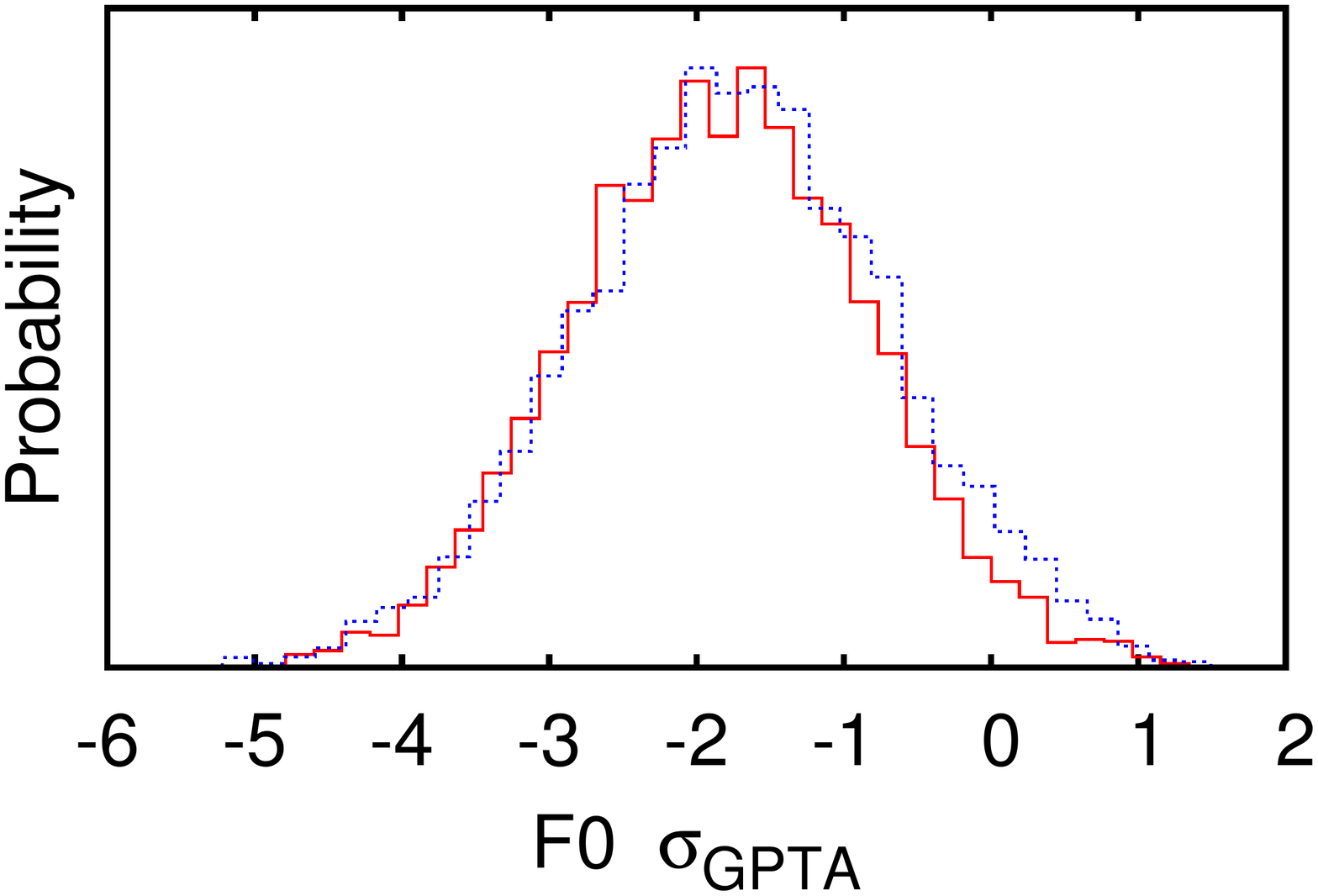} &
%
\includegraphics[width=50mm]{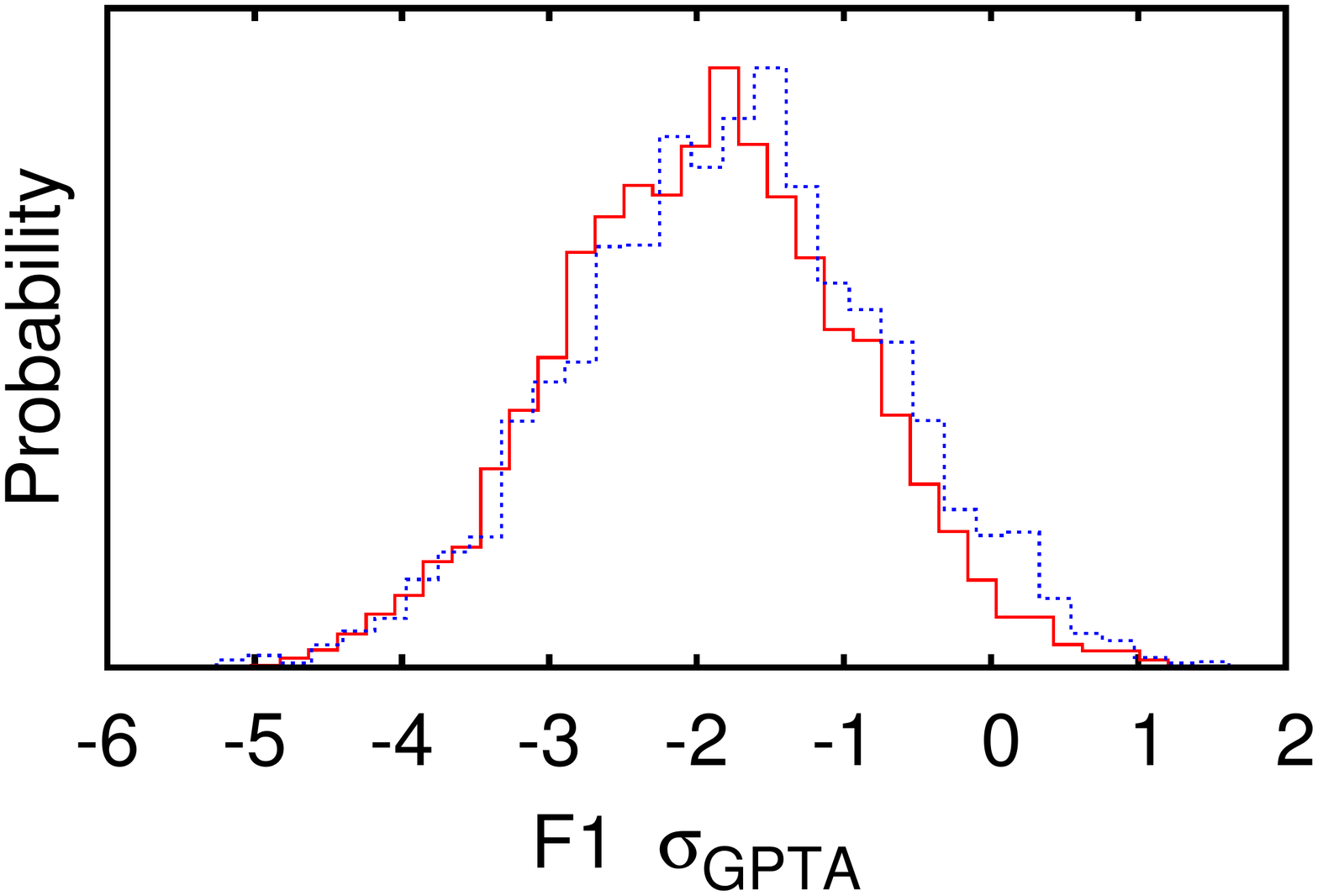} &
\includegraphics[width=50mm]{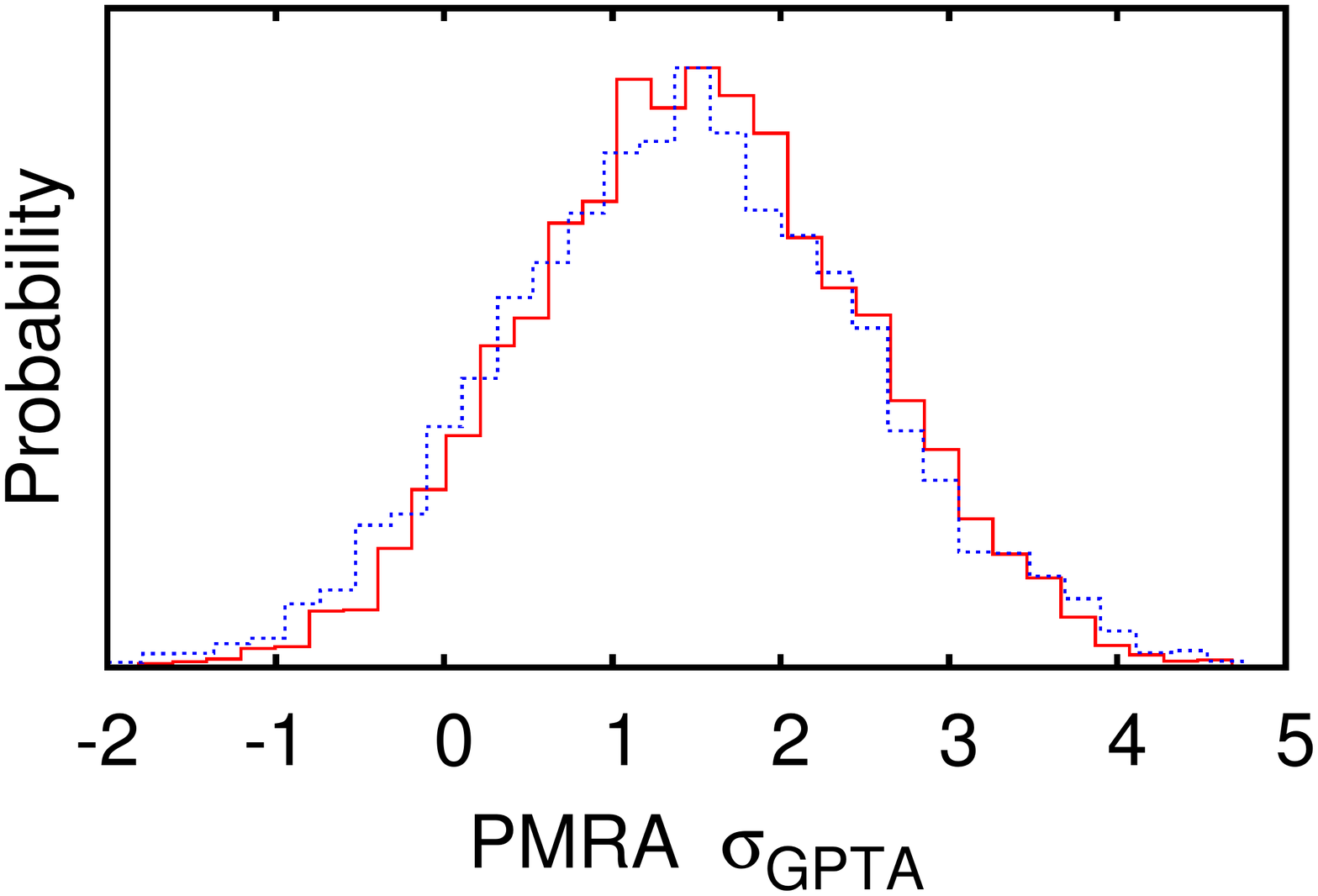} \\
\includegraphics[width=50mm]{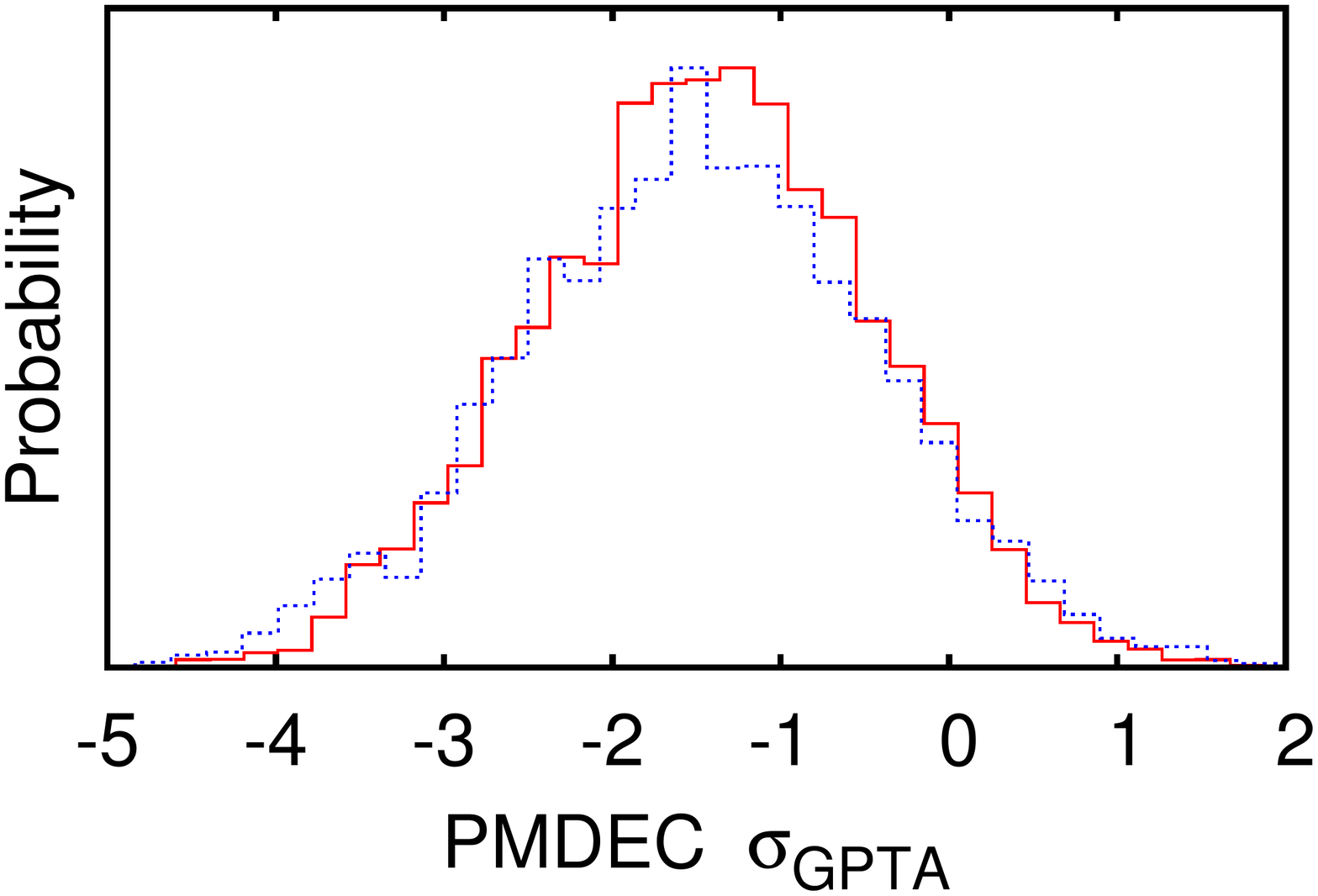} &
\includegraphics[width=50mm]{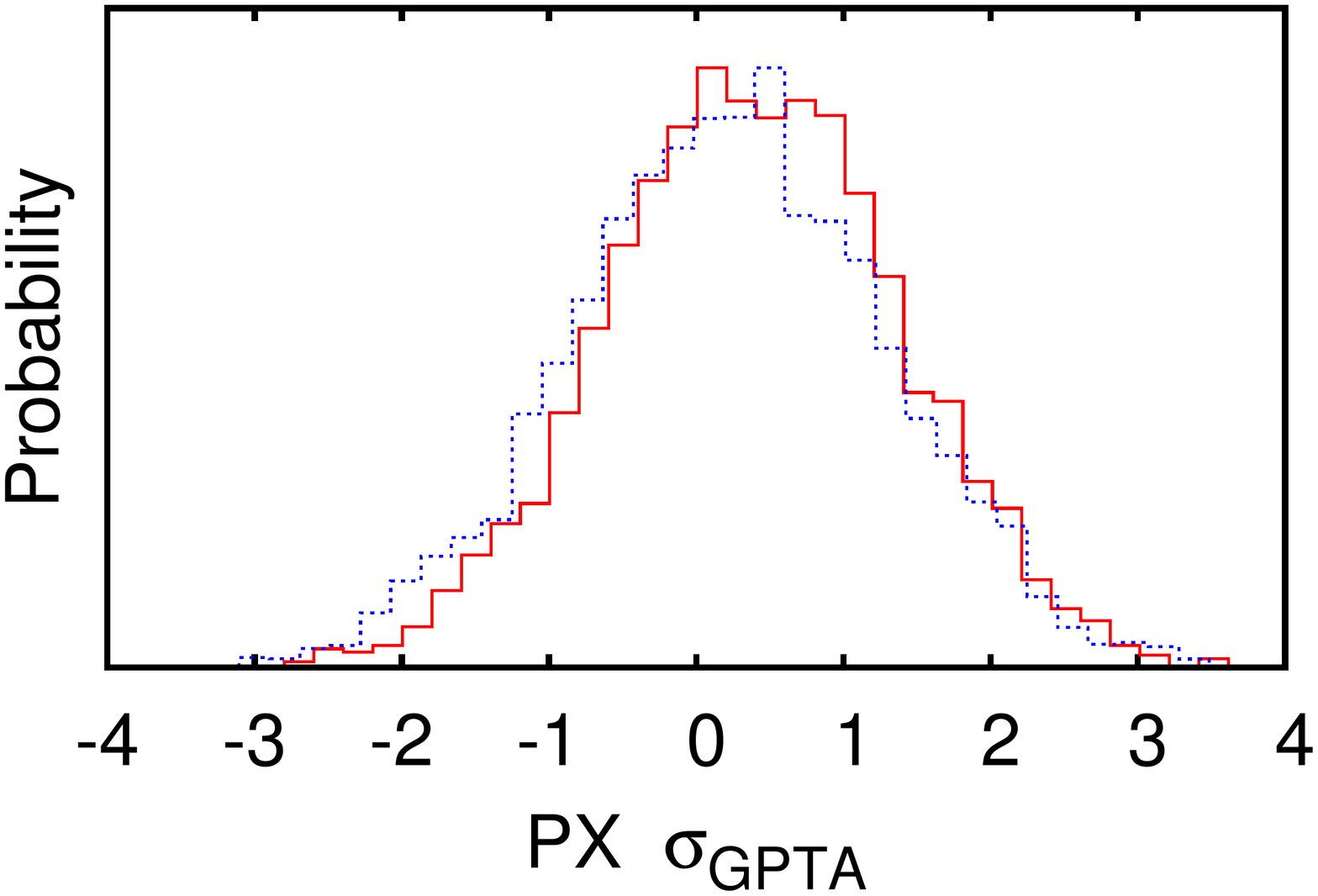} &

\end{array}$
\end{center}
\vspace{-0.8cm}
\caption{1-dimensional marginalised posteriors for the timing model in simulated dataset 1 for the isolated pulsar PSR J0030+0451 when performing the analysis using GPTA (red solid lines) and using TempoNest (blue dotted lines).  Values on the $x$-axes are given in terms of the $1\sigma$ uncertainties returned by the GPTA analysis, with the injected parameter value at 0 in all cases.  For all timing model parameters the two methods are completely consistent, both in terms of the parameter estimates and the uncertainties.  \label{figure:Sim1Comp}}
\end{figure*}

Fig \ref{figure:Sim1Comp} shows the one dimensional marginalised posteriors for the timing model in simulated dataset 1 using GPTA (green) and using TempoNest (red).  Values on the $x$-axes are given in terms of the $1\sigma$ uncertainties returned by the TempoNest analysis, with the injected parameter value at 0 in all cases.   Table \ref{Table:Sim1} lists the mean posterior values and associated $1\sigma$ uncertainties for all model parameters, including the profile parameters returned by the GPTA analysis.  For all timing model parameters the two methods are completely consistent, both in terms of the parameter estimates and the uncertainties as expected.

\subsection{Simulation 2}

\begin{figure*}
\begin{center}$
\begin{array}{c}
%
\includegraphics[width=120mm]{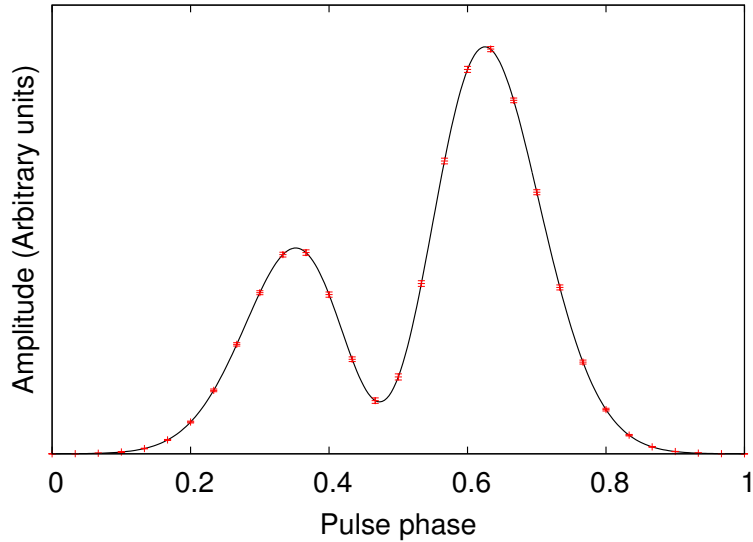} \\
\end{array}$
\end{center}
\vspace{-0.8cm}
\caption{Model pulse profile used in Simulation 2 (black line) and constraints placed on the profile at sampled intervals using GPTA.  For clarity we have increased the size of the uncertainties returned by the analysis by a factor of 100. \label{figure:Sim2prof}}
\end{figure*}

We now consider the case where there is significant red timing noise present in the data.  Here we use a more complex pulse profile generated using the lowest 3 shapelet coefficients described in Section \ref{section:Shapelets} which we show in Fig. \ref{figure:Sim2prof} (black line).  In order to produce constraints on the profile we store the value of the profile at each of the the 31 sampled points from every likelihood calculation and use these to construct posterior distributions for the value of the profile at those points.  The $1\sigma$ confidence intervals returned by this analysis are also shown in Fig. \ref{figure:Sim2prof}, however for clarity we have increased their size by a factor of 100.  The profile returned by the GPTA analysis is consistent within Gaussian statistics with the injected profile at all points.  We note that, while our model for the pulse profile is generated using the same set of basis functions as we then fit for, as shapelets form a complete set any profile can be reproduced using the shapelet formalism, and so this has no impact on the generality of the result.

\begin{figure*}
\begin{center}$
\begin{array}{cc}
\hspace{-1cm}
\includegraphics[width=100mm]{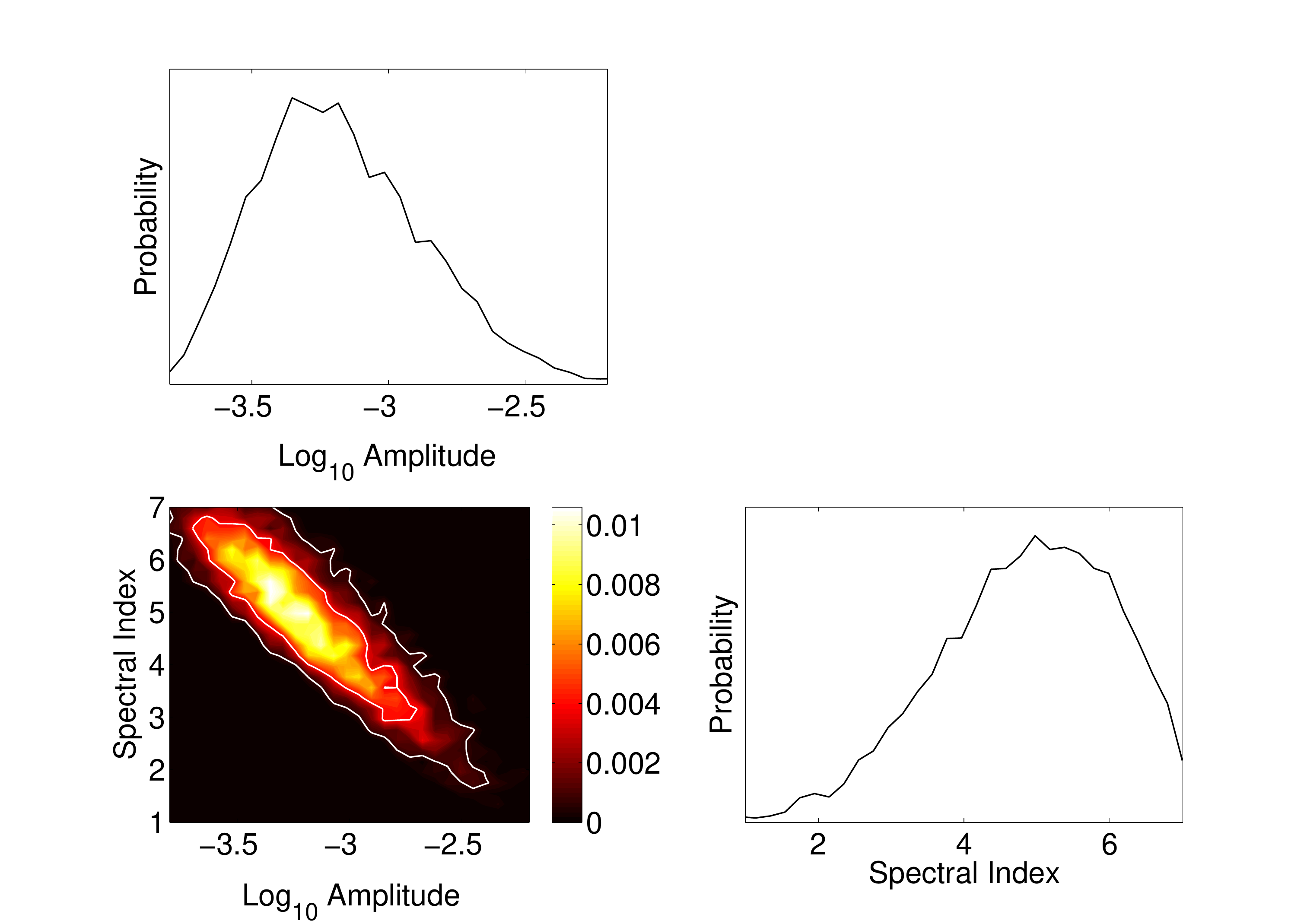} &
\hspace{-1cm}
\includegraphics[width=100mm]{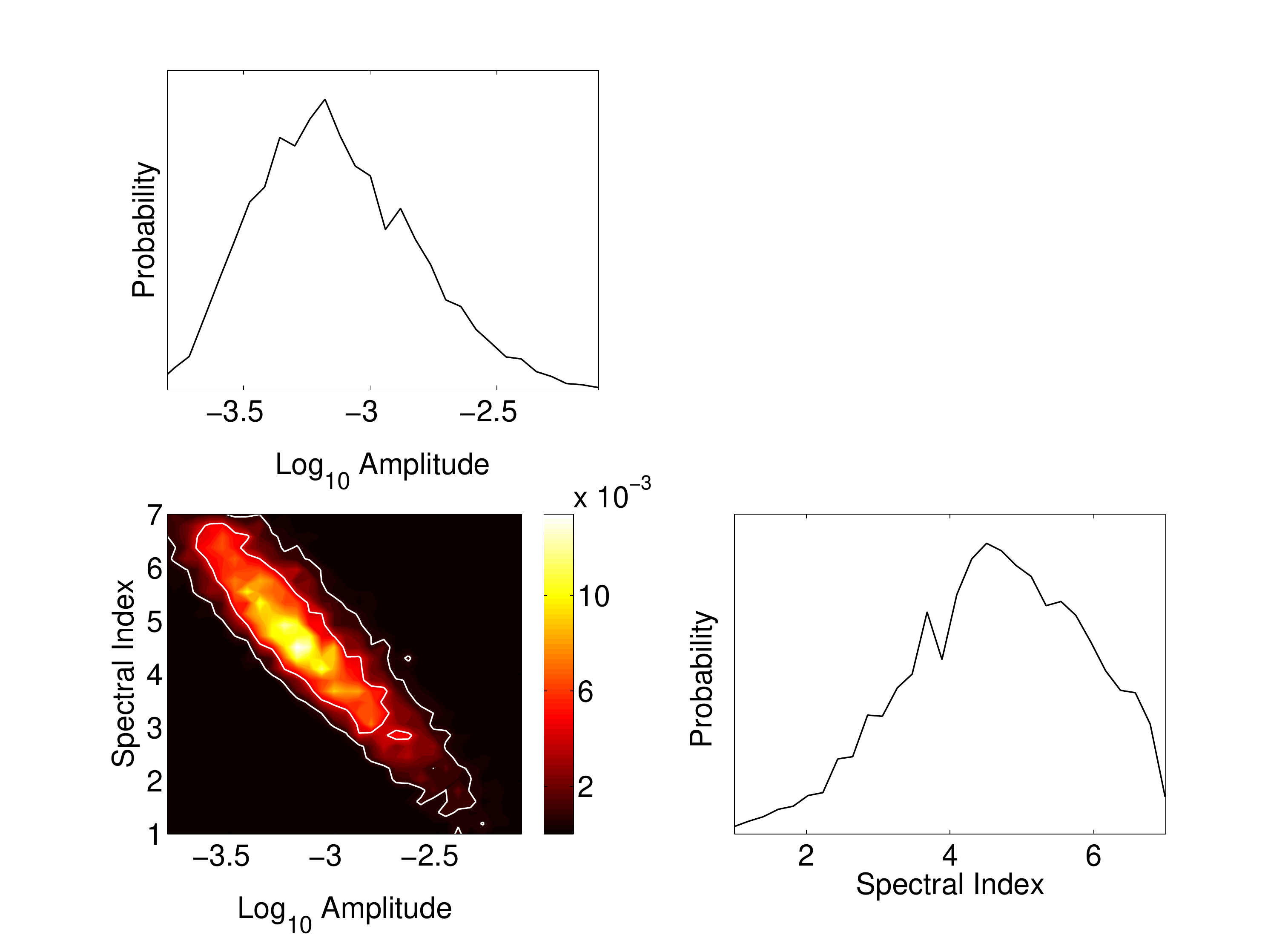} \\
\end{array}$
\end{center}
\caption{1 and 2-dimensional marginalised posteriors for the spectral index and amplitude of the injected red noise signal in simulated dataset 2 for the isolated pulsar PSR J0030+0451 when performing the analysis using GPTA (left) and using TempoNest (right). The scale in the 2-dimensional plot represents probability in both cases.  For all timing model and stochastic parameters the two methods are completely consistent, both in terms of the parameter estimates and the uncertainties.  \label{figure:Sim2red}}
\end{figure*}

As with Simulation 1 we construct a set of TOAs using the injected pulse profile and compare our analysis using GPTA to that of using TempoNest on the TOAs.  Table \ref{Table:Sim1} lists the mean posterior values and associated $1\sigma$ uncertainties for all timing model and stochastic parameters using these two methods, and in Fig. \ref{figure:Sim2red} we show the one and two-dimensional marginalised posteriors for the spectral index and amplitude of the injected red noise signal.  As in Simulation 1, our assumptions about the noise in the pulse profiles are correct in both cases, and so we find that our results for all parameters are consistent for the two methods.

\subsection{Simulation 3}
\label{Section:Sim4}

\begin{figure*}
\begin{center}$
\begin{array}{c}
%
\includegraphics[width=120mm]{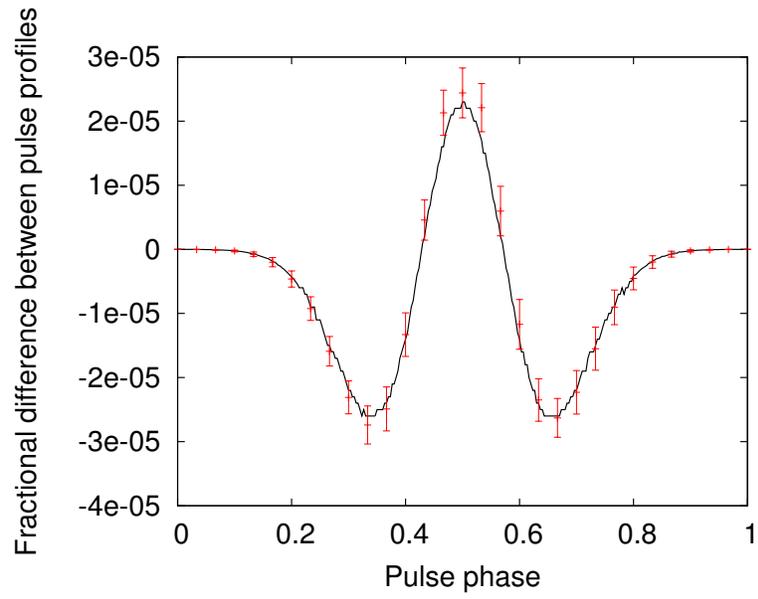} \\
\end{array}$
\end{center}
\vspace{-0.8cm}
\caption{Fractional difference between the two model pulse profiles used in Simulation 3 (black line) and constraints placed on this difference at sampled intervals using GPTA when including two model profiles in the analysis.\label{figure:Sim4prof}}
\end{figure*}

For the final simulation we consider the case where there are multiple observing frequencies in the dataset, and the pulse profile shows evolution from one frequency to the next.  We use different model profiles for each frequency where the fractional difference between is shown in Fig. \ref{figure:Sim4prof} (black line), alternating between them for each TOA.  Such division is completely arbitrary however, we could have just as easily considered evolution in time, and used the different pulse profiles for the first and second half of the dataset.

Table \ref{Table:Sim1} lists the mean posterior values and associated $1\sigma$ uncertainties for the timing model parameters when fitting for either an independent profile model for each frequency, or for a single average profile across both frequencies together. In Fig. \ref{figure:Sim4prof} we then show the constraints placed on the difference between the two profiles (blue points) when fitting for the two independent profiles in our analysis.

We find values for the $\log$ Evidence of 63252 and 63275 for models including one or two independent profiles respectively, giving a difference in the $\log$ evidence of 23 in favour of the two profile model, and thus provides unequivocal support for the additional parameters.  Despite this we find the timing model parameters for the one and two profile models to be completely consistent, both in terms of their parameter estimates and uncertainties.  We note however that in this case the difference between the two profiles was both relatively minor ($\sim$ 1 part in $10^5$), and symmetric, and so the effect the evolution has on the timing model estimation was negligible. 

\section{Conclusion}
\label{Section:Conclusions}

In this paper we have introduced a Bayesian framework that allows for the simultaneous analysis of the pulsar timing model, additional stochastic signals such as red `spin' noise and dispersion measure variations, in addition to a model for the pulse profile.  By using the timing model as a prior on the arrival time of the pulses, we can generate model TOAs at which to evaluate this profile and thereby provide a means of robustly estimating evolution in the profile in either frequency or in time. 

We demonstrated this method in a series of simulations and showed that when simulating folded profile data in the presence of only uncorrelated Gaussian noise, the timing parameters estimated using GPTA are completely consistent with those obtained using TempoNest when forming the TOAs in the traditional manner using the injected model profile as a template.  We then showed that GPTA is capable of robustly performing model selection between different evolutionary scenarios using the Bayesian evidence.

While the ultimate goal for Bayesian pulsar timing analysis should be to use the unfolded, individual pulses, the increased volume in data this represents currently makes it  computationally intractable to do so.  The folded data, however, presents a much less significant computational challenge.  As GPTA requires no large, dense matrix operations, even when including long-term variations in the TOAs, or correlated signals in the profile data, it can be used on current pulsar timing datasets, on typical workstation computers.  

The simplicity of our method allows for many possible extensions.   In this paper we discussed two such modifications, including models for correlated noise in the pulse profiles, and pulse broadening due to scattering in the interstellar medium, however additional physical effects that modify the pulse profile could also be incorporated into this analysis pipeline with little extra effort.  As such, while it is a marked departure from standard pulsar timing analysis methods, it has clear applications for both current and new datasets, such as those from the European Pulsar Timing Array (EPTA, \citep{2008AIPC..983..633J}) and International Pulsar Timing Array (IPTA, \cite{2013CQGra..30v4010M}).

\section{Acknowledgements}

This research is the result of the common effort to directly detect gravitational waves using pulsar timing, known as the European Pulsar Timing Array (EPTA)  \footnote{www.epta.eu.org/}.

\label{lastpage}


\begin{thebibliography}{}
\setlength{\labelwidth}{0pt} 

\bibitem[\protect\citeauthoryear{AMI Consortium et 
al.}{2012}]{2012arXiv1210.7771C} AMI Consortium, et al., 2012, arXiv, 
arXiv:1210.7771 

\bibitem[\protect\citeauthoryear{Berry, Hobson, 
\& Withington}{2004}]{2004MNRAS.354..199B} Berry R.~H., Hobson M.~P., Withington S., 2004, MNRAS, 354, 199


\bibitem[\protect\citeauthoryear{Bhat et al.}{2004}]{2004ApJ...605..759B} 
Bhat N.~D.~R., Cordes J.~M., Camilo F., Nice D.~J., Lorimer D.~R., 2004, 
ApJ, 605, 759  

\bibitem[\protect\citeauthoryear{Bhat et 
al.}{2007}]{2007A&A...462..257B} Bhat N.~D.~R., Gupta Y., Kramer M., Karastergiou A., Lyne A.~G., Johnston S., 2007, A\&A, 462, 257 


\bibitem[\protect\citeauthoryear{Cronyn}{1970}]{1970Sci...168.1453C} Cronyn 
W.~M., 1970, Sci, 168, 1453 



\bibitem[\protect\citeauthoryear{Demorest et 
al.}{2013}]{2013ApJ...762...94D} Demorest P.~B., et al., 2013, ApJ, 762, 94 

\bibitem[\protect\citeauthoryear{Feroz 
\& Hobson}{2008}]{2008MNRAS.384..449F} Feroz F., Hobson M.~P., 2008, MNRAS, 384, 449 


\bibitem[\protect\citeauthoryear{Feroz, Hobson, 
\& Bridges}{2009}]{2009MNRAS.398.1601F} Feroz F., Hobson M.~P., Bridges M., 2009, MNRAS, 398, 1601 

\bibitem[\protect\citeauthoryear{Hankins 
\& Cordes}{1981}]{1981ApJ...249..241H} Hankins T.~H., Cordes J.~M., 1981, ApJ, 249, 241 

\bibitem[\protect\citeauthoryear{Hobson, Bridle, 
\& Lahav}{2002}]{2002MNRAS.335..377H} Hobson M.~P., Bridle S.~L., Lahav O., 2002, MNRAS, 335, 377 


\bibitem[\protect\citeauthoryear{Janssen et 
al.}{2008}]{2008AIPC..983..633J} Janssen G. H., Stappers B. W., Kramer M., Purver M., Jessner A., Cognard
I., 2008, in Bassa C., Wang Z., Cumming A., Kaspi V. M., eds, AIP Conf.
Proc. Vol. 983, 40 Years of Pulsars: Millisecond Pulsars, Magnetars and
More. Am. Inst. Phys., New York, p. 633


\bibitem[\protect\citeauthoryear{Kaspi, Taylor, 
\& Ryba}{1994}]{1994ApJ...428..713K} Kaspi V.~M., Taylor J.~H., Ryba M.~F., 1994, ApJ, 428, 713 


\bibitem[\protect\citeauthoryear{Kelly 
\& McKay}{2004}]{2004AJ....127..625K} Kelly B.~C., McKay T.~A., 2004, AJ, 127, 625 

\bibitem[\protect\citeauthoryear{Lee et al.}{2012}]{2012MNRAS.423.2642L} 
Lee K.~J., Bassa C.~G., Janssen G.~H., Karuppusamy R., Kramer M., Smits R., 
Stappers B.~W., 2012, MNRAS, 423, 2642 


\bibitem[\protect\citeauthoryear{Lentati et 
al.}{2013}]{2013MNRAS.430.2454L} Lentati L., et al., 2013, MNRAS, 430, 2454 


\bibitem[\protect\citeauthoryear{Lentati et 
al.}{2013}]{2013PhRvD..87j4021L} Lentati L., Alexander P., Hobson M.~P., 
Taylor S., Gair J., Balan S.~T., van Haasteren R., 2013,  Phys. Rev. D, 87, 104021 


\bibitem[\protect\citeauthoryear{Lentati et 
al.}{2013}]{2013arXiv1310.2120L} Lentati L., Alexander P., Hobson M.~P., 
Feroz F., van Haasteren R., Lee K., Shannon R.~M., 2013, arXiv, 
arXiv:1310.2120 

\bibitem[\protect\citeauthoryear{Lentati, Hobson, 
\& Alexander}{2014}]{2014arXiv1405.2460L} Lentati L., Hobson M.~P., Alexander P., 2014, arXiv, arXiv:1405.2460 


\bibitem[\protect\citeauthoryear{Lyne 
\& Manchester}{1988}]{1988MNRAS.234..477L} Lyne A.~G., Manchester R.~N., 1988, MNRAS, 234, 477 

\bibitem[\protect\citeauthoryear{Manchester et 
al.}{2005}]{2005AJ....129.1993M} Manchester R.~N., Hobbs G.~B., Teoh A., 
Hobbs M., 2005, AJ, 129, 1993 

\bibitem[\protect\citeauthoryear{Manchester 
\& IPTA}{2013}]{2013CQGra..30v4010M} Manchester R.~N., IPTA, 2013, CQGra, 30, 224010 



\bibitem[\protect\citeauthoryear{Matsakis, Taylor, 
\& Eubanks}{1997}]{1997A&A...326..924M} Matsakis D.~N., Taylor J.~H., Eubanks T.~M., 1997, A\&A, 326, 924 


\bibitem[\protect\citeauthoryear{O'Ruanaidh \& Fitzgerald}{1996}]{Thermo} O'RuanaidhJ. J. K., Fitzgerald W. J., 1996, Springer-Verlag, New York


\bibitem[\protect\citeauthoryear{Refregier}{2003}]{2003MNRAS.338...35R} 
Refregier A., 2003, MNRAS, 338, 35 


\bibitem[\protect\citeauthoryear{Refregier 
\& Bacon}{2003}]{2003MNRAS.338...48R} Refregier A., Bacon D., 2003, MNRAS, 338, 48 


\bibitem[\protect\citeauthoryear{Rickett}{1990}]{1990ARA&A..28..561R} Rickett B.~J., 1990, ARA\&A, 28, 561 

\bibitem[\protect\citeauthoryear{Shannon et 
al.}{2014}]{2014MNRAS.443.1463S} Shannon R.~M., et al., 2014, MNRAS, 443, 
1463 

\bibitem[\protect\citeauthoryear{Skilling}{2004}]{2004AIPC..735..395S} 
Skilling J., 2004, in Fischer R., Preuss R., von Toussaint U., eds, AIP Conf.
Proc. Vol. 735, Bayesian Inference and Maximum Entropy Methods in
Science and Engineering. Am. Inst. Phys., New York, p. 395






\end{thebibliography}
\end{document}